\documentclass[preprint]{aastex}

\usepackage{amsmath,amssymb,bm,mathrsfs}

\DeclareFontFamily{OMS}{bcmsy}{}
\DeclareFontShape{OMS}{bcmsy}{m}{n}{%
	<5> bcmsy5
	<6> bcmsy6
	<7> bcmsy7
	<8> bcmsy8
	<9> bcmsy9
	<10-> bcmsy10}{}
\DeclareSymbolFont{symbols}{OMS}{bcmsy}{m}{n}

\DeclareFixedFont{\elevensy}{OMS}{bcmsy}{m}{n}{10.95pt}
\DeclareFixedFont{\ninesy}{OMS}{bcmsy}{m}{n}{9pt}

\newcommand{\<}{{\kern-5pt}}

\newcommand{\thrj}[6]{\biggl(
	\arraycolsep .2em
	\begin{matrix}
	#1&#2&#3\\
	#4&#5&#6\\
	\end{matrix}\biggr)}
\newcommand{\sixj}[6]{\biggl\{
	\arraycolsep .2em
	\begin{matrix}
	#1&#2&#3\\
	#4&#5&#6\\
	\end{matrix}\biggr\}}
\newcommand{\ninj}[9]{\left\{
	\arraycolsep .2em
	\renewcommand{\arraystretch}{.9}
	\begin{matrix}
	#1&#2&#3\\
	#4&#5&#6\\
	#7&#8&#9\\
	\end{matrix}\right\}}

\def\mlangle{\kern.175em\langle}	% remove for TeX
\def\mrangle{\rangle\kern.175em}	% remove for TeX
\def\ket#1{|{#1}\mrangle}
\def\bra#1{\mlangle{#1}|}
\def\redmat#1#2#3{\bra{#1}\kern -1pt|#2|\kern -1pt\ket{#3}}

\newcommand{\apx}[1]{^{\mbox{\tiny{(#1)}}}}

\begin{document}

\title{Frequency Redistribution Function for the Polarized Two-Term Atom}

\author{R.\ Casini,$^a$
M.\ Landi Degl'Innocenti,$^b$ R.\ Manso Sainz,$^c$ E.\ Landi
Degl'Innocenti,$^d$ M.\ Landolfi$^e$
}
%\author{E. Landi Degl'Innocenti$^d$}

\affil{$^a$High Altitude Observatory, National Center for Atmospheric
Research,\footnote{The National Center for Atmospheric Research is sponsored
by the National Science Foundation.}\break
P.O.~Box 3000, Boulder, CO 80307-3000, U.S.A.}
\affil{$^b$Istituto Nazionale di Astrofisica, Largo E.~Fermi 5, I-50125
Firenze, Italy}
\affil{$^c$Instituto de Astrof\'{\i}sica de Canarias, c/ V\'{\i}a
L\'actea s/n, E-38205 La Laguna, Tenerife, Spain}
\affil{$^d$Dipartimento di Astronomia e Scienze dello Spazio, Universit\`a
di Firenze, Largo E.~Fermi 2, I-50125 Firenze, Italy} 
\affil{$^e$Osservatorio Astrofisico di Arcetri, Largo E.~Fermi 5, I-50125
Firenze, Italy}

\begin{abstract}
We present a generalized frequency redistribution function for the
polarized two-term atom in an arbitrary magnetic field. This result is
derived within a new formulation of the quantum problem of coherent
scattering of polarized radiation by atoms in the collisionless
regime. The general theory, which is based on a diagrammatic treatment
of the atom-photon interaction, is still work in progress. However,
the results anticipated here are relevant enough for the study
of the magnetism of the solar chromosphere and of interest for 
astrophysics in general.
\end{abstract}

\section{Introduction}
The diagnosis of magnetic fields in astrophysical plasmas relies on the
measurement and interpretation of the polarization signature of the
magnetic fields in spectral lines. The physical conditions of these
plasmas vary greatly depending on the type of astronomical objects
considered. As a consequence a variety of simultaneous, and sometimes
competing, effects -- radiative, collisional, and from plasma electric
and magnetic fields -- need to be taken into consideration in the 
modeling of the observed radiation.

In dense plasmas, such as those characteristic of stellar
atmospheres, strong gradients of temperature and density, as well as of
magnetic and velocity fields, often occur (e.g., due to atmosphere 
stratification under the effect of gravitation and pressure gradients, 
or because of the presence of convective motions and dynamo actions), 
which need to be taken into account in modeling the 
transport of polarized radiation through the plasma. This is certainly 
the case for the strongest absorption lines of the solar spectrum,
giving rise to a great diversity of line profile shapes and degrees of
polarization, especially when observed with high temporal and spatial
resolution \cite[e.g.,][]{BL09}

Besides these non-local, radiative transfer effects, the particular
excitation conditions of a magnetized plasma away from local 
thermodynamic equilibrium often represents a challenging problem for
the description of the interaction between matter and polarized
radiation. Apart from the well-studied 
Zeeman effect, subtle quantum
processes associated with the polarized state of the atoms are often at
play in stellar atmospheres, such as the Hanle effect, or other
quantum effects associated with level-crossing physics. While these
processes offer an opportunity for a highly refined diagnostics of 
magnetized plasmas, they also present theoretical and computational 
difficulties that have commonly prevented their full exploitation 
in the astrophysical context.

Modeling of partial redistribution of the photon's frequency in
the scattering of polarized radiation is perhaps one of the most 
difficult problems to treat at the fundamental level, as this generally
requires to describe the interaction of radiation with matter beyond the 
lowest order of approximation, which is instead adequate in the case
of absorption and emission lines that are formed under conditions of
complete frequency redistribution. For example, when isotropic
plasma collisions are effective at destroying the coherence of 
multi-photon interactions, the scattering of radiation can 
safely be described as the incoherent succession of single-photon
absorption and emission. It can also be shown that irradiation of the 
atom by a spectrally flat radiation allows a description of the 
scattering process that is phenomenologically identical to that 
provided by such two-step,
single-photon processes of absorption and incoherent re-emission
(e.g., \citealt{He54}, \S20; see also \citealt{Sa67}, \S2.6, and
Section~\ref{sec:balance} of this paper). 
Therefore, the regime 
for which the frequency coherence effects characteristic of radiation 
scattering are instead important is typically one of low plasma densities 
(i.e., low collisional rates) and highly spectrally modulated radiation, such
as that emerging from stellar atmospheres in correspondence of very deep 
absorption lines.

This paper introduces a general frequency redistribution function for
the polarized two-term atom, and then considers particular cases of it,
which are relevant to the investigation of strong resonance lines in 
the solar spectrum, for which partial redistribution effects are deemed 
to be 
important. However, the results presented here are also valid in the case 
of subordinate lines, when both the upper and lower terms are
radiatively broadened by spontaneous de-excitation. This study is based 
on a novel formalism \cite[][work in progress]{Ca15} that, within the
framework of non-relativistic Quantum Electrodynamics, describes the 
time evolution of the atomic system and the radiation field in terms of 
\emph{propagators}. 
This formalism relies on a diagrammatic representation of the 
atom-photon interaction, which allows to correctly identify and enumerate 
the types of processes that contribute to the scattering of polarized
radiation. At the lowest order of interaction, it reproduces 
the theory of polarized line formation for complete redistribution of 
frequency \citep{LL04}.

Over the past four decades, several authors have addressed the 
problem of partial redistribution in polarized spectral lines. We will 
not provide a detailed account of the progress in this area, but we 
limit ourselves to simply tracing the main lines of work. 
Notable are the works of \cite{Om72,Om73}, and of Heinzel and
collaborators \cite[e.g.,][]{He81,HH82,Hu82}, all of which derive 
essentially from the seminal work of \cite{FV62}, who had extended
the impact theory of pressure broadening developed by \citeauthor{An49} 
(\citeyear{An49}; see also \citealt{Ba58}) to second order, in order to 
study the line shape of molecular Raman scattering. 
\cite{FV62} assumed the diagonality of the density matrix of the 
initial (lower) state, as did the subsequent works that were based on 
the same formalism. This assumption corresponds to 
the hypothesis of non-coherent lower term (see Sect~\ref{sec:special}).
More recently, \cite{St94} followed a heuristic approach built upon the 
Kramers-Heisenberg scattering amplitude in order to derive a 
semi-classical theory of partial redistribution, which has 
enabled the derivation of redistribution functions similar to those 
presented in this paper \cite[see, e.g.,][]{Sm13},
again for atoms with non-coherent lower term.
% and has infinitely sharp energy levels. 

Other works have approached the general problem of spectral line
formation, investigating frequency redistribution effects in the
statistical equilibrium of the atom and in the transport of polarized
radiation through an absorptive and scattering medium.
\cite{LT71} followed an approach based on \cite{He54}, which allowed for 
the presence of atomic coherence (i.e., non-diagonal density matrix) in 
the lower term.
Remarkably their work led to results for the scattering redistribution 
function that are in full agreement with ours.
\cite{Co82} investigated the redistribution problem relying on 
the quantum-regression theorem \cite[e.g.,][]{Lo73}, and derived 
results that agree with those of \cite{Om72}.
\cite{La97} extended the first-order theory of polarized line formation 
of \cite{La83} to include partial redistribution effects in a two-term
atom. This was done by describing the atomic states in terms of energy 
\emph{metalevels} \citep[or energy \emph{sub-states};][]{WS53},
and proposing a heuristic generalization of the atomic density 
matrix based upon the ``metalevel'' idea. Their formalism naturally 
reproduces the results of the first-order theory of polarized 
line formation, in the limit of complete redistribution (e.g., for a 
flat-spectrum illumination of the atom), and it was applied successfully 
to the problem of the scattering polarization in the \ion{Na}{1} D-doublet
\citep{La98}. 
Finally, \cite{Bo97a,Bo97b} formally extended the line formation
theory of \cite{La83} to higher orders of perturbation, at the same time
relying also on the results of \cite{Ba58} to describe relaxation
effects on the atomic system from the interaction with a thermal bath of
colliding perturbers. The redistribution function for the
two-level atom with unpolarized lower level derived from that theory was 
shown to be in agreement with the results of \cite{Om72,Om73}, 
and of the heuristic approaches of \cite{St94} and \cite{La97} in the 
absence of collisions.

In this paper, we introduce in Section~\ref{sec:theory} some fundamental 
theoretical 
results that lie at the basis of our treatment of partial frequency 
redistribution of polarized radiation. In Section~\ref{sec:special} we 
re-derive several special forms of the
redistribution function known from the past literature, and also 
present a generalization of those previous results that allows for
atomic coherence in the lower term. In Section~\ref{sec:lifetimes} 
we briefly discuss the definition of the radiative lifetimes of 
atomic states perturbed by the presence of external fields. This is an 
important clarification for the applicability of the results here
presented. (As usual, collisional widths can be added to the natural
widths of the atomic level in a phenomenological way, in the limit of 
the impact approximation. See, e.g., \citealt{An49,LT71}.)
In Section~\ref{sec:application} we specialize the equation of
radiative transfer with partial redistribution to the general case of a
polarized two-term atom with hyperfine structure. The cases of simpler
atomic structures can be obtained from this more general case, and as an
illustration we re-derive the expression of the scattering emissivity for
the two-term atom with unpolarized lower levels and in the limit 
of zero magnetic field, which has been derived previously through the metalevel 
approach \citep{La97}. Finally, in
Section~\ref{sec:balance} we verify that the generalized 
radiative transfer equation with the inclusion of coherent scattering
satisfies the fundamental condition that the energy flux of the
radiation through 
a closed surface containing the scatterer must be zero, in the absence
of collisions.

\section{Theory} \label{sec:theory}
We consider a two-term atom with upper levels $\{u,u',u'',\ldots\}$ 
and lower levels $\{l,l',l'',\ldots\}$. These levels represent the
energy eigenstates of the atomic Hamiltonian in the presence of the
external fields. In general, we allow for these levels to be 
arbitrarily polarized, and therefore we assume the existence of 
atomic coherence among the levels, described by the (complex)
non-diagonal elements of the atomic density matrix, such as 
$\rho_{uu'}$ and $\rho_{ll'}$. We want to study the scattering of 
polarized radiation 
by an ensemble of such atoms, the radiation being described by
an arbitrary incident beam of wave vector $\bm k$ and polarization 
state $(\lambda,\mu)$, and an emerging beam of given wave vector 
$\bm k'$ and polarization state $(\lambda',\mu')$. In particular, we
want to focus on the \emph{coherent} part of the evolution equation of 
the radiation field, where the upper levels $u,u'$ enter only as 
virtual states. This part only appears when we push the development of
the formal theory to include multi-photon effects \cite[e.g.,][]{Bo97a}.

We thus find that the equation for the transport of polarized 
radiation \emph{for the two-term atom}, expressed in the atomic 
reference frame, can be written in the form
\begin{equation} \label{eq:RT}
\frac{d}{dt}\,I_{\lambda'\mu'}(\bm{k'})
=
	-\sum_{\lambda\mu}\kappa_{\lambda'\mu'}^{\lambda\mu}(\bm{k'})\,
	I_{\lambda\mu}(\bm{k'})
	+\varepsilon\apx{1}_{\lambda'\mu'}(\bm{k'})
	+\varepsilon\apx{2}_{\lambda'\mu'}(\bm{k'})\;.
\end{equation}
The first-order terms, $\varepsilon\apx{1}_{\lambda'\mu'}(\bm{k'})$ and
$\kappa_{\lambda'\mu'}^{\lambda\mu}(\bm{k'})$, whose explicit expressions 
we omit for the moment (see Sect.~\ref{sec:application}),
describe, respectively, the emission of radiation 
of wave vector $\bm k'$ and polarization state $(\lambda',\mu')$ due to 
the spontaneous de-excitation of the atom from all possible states 
$\rho_{uu'}$, and the absorption rate of the
incident radiation propagating \emph{along the same vector} $\bm k'$ 
(but in all possible polarization states) when the atom is in any 
of the possible states $\rho_{ll'}$. We neglect, 
here and in the following, any contribution of
stimulated emission to the expression of 
$\kappa_{\lambda'\mu'}^{\lambda\mu}(\bm{k'})$. We make this choice in 
order to facilitate comparison of our results with those in the former 
literature on the subject of partial redistribution.
The second-order term $\varepsilon\apx{2}_{\lambda'\mu'}(\bm{k'})$
describes instead the scattering of radiation by 
the atom from all possible lower states $\rho_{ll'}$, via all 
possible intermediate (virtual) levels $u,u'$, and final levels $l''$.

Thus, the first-order terms in Equation~(\ref{eq:RT}) account for the 
transfer of radiation along the direction $\bm k'$, while the
second-order term describes the phenomenon of coherent scattering. 
We note that we use here the term ``coherent'' in the broader sense of 
``memory preserving'', rather than in the usual restricted sense of 
``frequency preserving''. This seems to be a better interpretation in 
light of the theoretical development that led to Equation~(\ref{eq:RT})
\citep[][work in progress]{Ca15}.

%In the following, we consider in detail the coherent-scattering term,
%which can explicitly be written in the following form,
The coherent-scattering term has the following expression,
\begin{eqnarray} \label{eq:2emiss}
\varepsilon\apx{2}_{\lambda'\mu'}(\bm{k'})
&\equiv&
	\frac{1}{\hbar^4}\sum_{ll'}\rho_{ll'}\sum_{uu'l''}
	\sum_{\lambda\mu,\bm{k}}
	Q^\ast_{ul''}(\lambda',\bm{k'})
	Q_{u'l''}(\mu',\bm{k'})
	Q_{ul}(\lambda,\bm{k})
	Q_{u'l'}^\ast(\mu,\bm{k})\,
      I_{\lambda\mu}(\bm{k}) \nonumber \\
\noalign{\vskip -6pt}
&&\hphantom{+\frac{1}{\hbar^4}\sum_{ll'}\rho_{ll'}\sum_{uu'l''}
	\sum_{\lambda\mu,\bm{k}}} \times
	\Bigl( \Psi_{u'l',l''ul}^{-k,+k'-k} +
	\bar\Psi_{ul,l''u'l'}^{-k,+k'-k} \Bigr)\;.
%	{\cal R}(\Omega_u,\Omega_{u'};
%	\Omega_l,\Omega_{l'},\Omega_{l''};
%	\omega_k,\omega_{k'})\;. 
\end{eqnarray}
First we note that the sum over $\bm k$ actually stands for a 
continuous integration on both frequency and direction of propagation, 
so we can operate the usual substitution, which holds in the atomic
frame of rest,
\begin{equation} \label{eq:cont.rad}
\sum_{\bm{k}} I_{\lambda\mu}(\bm{k}) \longrightarrow
	\frac{1}{2\pi^2}\frac{{\cal V}}{c^3} 
	\int_0^\infty d\omega_k\,\omega_k^2 
	\oint \frac{d\bm{\hat{k}}}{4\pi}\,
	I_{\lambda\mu}(\bm{k})\;,
\end{equation}
where $\cal V$ is the volume of the quantization box. The ``vertex'' 
form factor, $Q_{ab}(\lambda,\bm{k})$, in the electric-dipole 
approximation, is given by \citep[e.g.,][]{LL04}
\begin{equation} \label{eq:rdipole}
Q_{ab}(\lambda,\bm{k})
	=\sqrt{\frac{2\pi e_0^2\hbar}{{\cal V}}\,
	\omega_k}\,
	\sum_q(-1)^q(r_q)_{ab}\,
	(\bm{e}_{\bm{\hat{k}}\lambda})_{-q}\;,
%Q_{ab}(\lambda,\bm{k}) Q_{cd}^\ast(\mu,\bm{k})
%	=\frac{2\pi e_0^2\hbar}{{\cal V}}\,\omega_k
%	\sum_{qq'}(-1)^{q+q'}(r_q)_{ab}(r_{q'})^\ast_{cd}\,
%	(\bm{e}_{\bm{\hat{k}}\lambda})_{-q}
%	(\bm{e}_{\bm{\hat{k}}\mu})_{-q'}^\ast\;,
\end{equation}
where we adopted the spherical-tensor representation of the
electric-dipole and polarization unit vectors, $\bm{r}$ and 
$\bm{e}_{\bm{\hat{k}}\alpha}$, respectively.

Finally, the complex line profiles $\Psi_{ab,cde}^{\pm h,\pm k\pm l}$
are given by
\begin{eqnarray} \label{eq:Psi}
\Psi_{ab,cde}^{\pm h,\pm k\pm l}
&\equiv&
\frac{-{\rm i}}{
(\omega_{ac}\pm\omega_h\mp\omega_l\mp\omega_k
	+{\rm i}\epsilon_a+{\rm i}\epsilon_c)
(\omega_{ad}\pm\omega_h\mp\omega_l+{\rm i}\epsilon_a+{\rm i}\epsilon_d)
(\omega_{ae}\pm\omega_h+{\rm i}\epsilon_a+{\rm i}\epsilon_e)
} \nonumber \\
&+&\frac{-{\rm i}}{
(\omega_{ac}\pm\omega_h\mp\omega_l\mp\omega_k
	+{\rm i}\epsilon_a+{\rm i}\epsilon_c)
(\omega_{bc}\mp\omega_l\mp\omega_k+{\rm i}\epsilon_b+{\rm i}\epsilon_c)
(\omega_{cd}\pm\omega_k-{\rm i}\epsilon_c+{\rm i}\epsilon_d)
} \nonumber \\
&-&\frac{-{\rm i}}{
(\omega_{ad}\pm\omega_h\mp\omega_l+{\rm i}\epsilon_a+{\rm i}\epsilon_d)
(\omega_{bd}\mp\omega_l+{\rm i}\epsilon_b+{\rm i}\epsilon_d)
(\omega_{cd}\pm\omega_k-{\rm i}\epsilon_c+{\rm i}\epsilon_d)
}\;.
\end{eqnarray}
For notational convenience, we introduced the Bohr frequency, 
$\omega_{pq}=\omega_p-\omega_q$, for any two levels $p$ and $q$. The
quantity $\epsilon_p$ represents the width of the level $p$. In
Equation~(\ref{eq:2emiss}), $\bar\Psi$ indicates complex conjugation of the
profile $\Psi$. 

\begin{figure}[t!]
\centering
\includegraphics[width=2.65in]{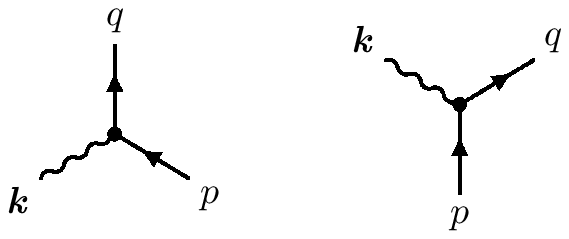}\vspace{0pt}
\includegraphics[width=\hsize]{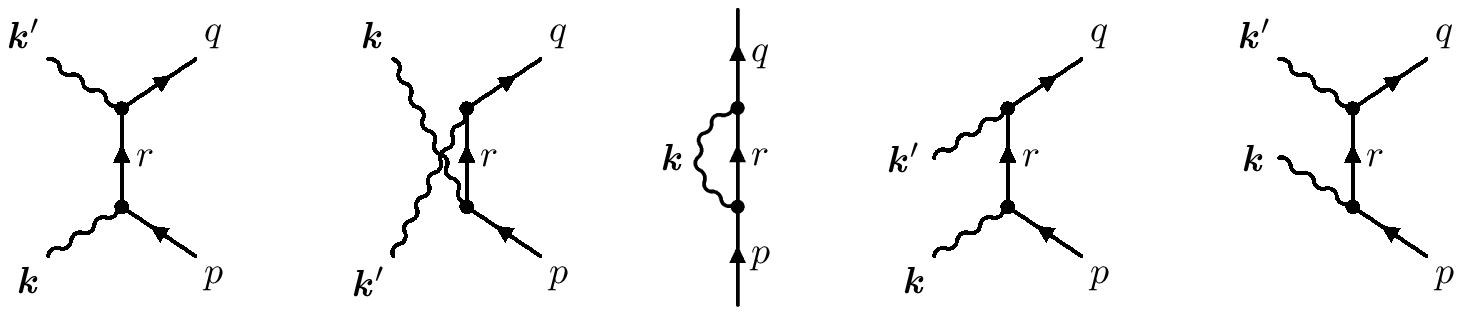}
\caption{\label{fig:diagrams}
Feynman diagrams representing the first-order (i.e., single photon) 
processes of absorption and emission (top row), and the second-order
(i.e., two-photon) processes describing scattering, two-photon 
absorption, and two-photon emission (bottom row). The central diagram 
in the bottom row represents the lowest-order radiative correction to 
the atomic propagator. Its main effect is that of ``dressing'' a given 
atomic state with a finite lifetime associated with spontaneous
de-excitation.}
\end{figure}

Both Equations~(\ref{eq:2emiss}) and (\ref{eq:Psi}) are derived from
first principles. The supporting theory, which is still work in
progress, follows a diagrammatic approach to the derivation of the 
evolution equation for a general atomic system interacting with a 
polarized radiation field.
This approach leads to the same perturbation series for the evolution 
equation of the coupled system atom+radiation as that of \cite{Bo97a},
and the initial conditions for the system's density matrix are also
handled in a similar way. Somewhat surprisingly, the fundamental 
results that we derive separately for the statistical equilibrium of 
the atomic system and for the transport of polarized radiation do not 
agree with the work of \cite{Bo97a,Bo97b}, although both approaches 
lead to the same redistribution function for the two-level atom with 
unpolarized lower level and in the absence of collisions.

It is worth spending a few words on the notation adopted for the
profiles $\Psi_{ab,cde}^{\pm h,\pm k\pm l}$. The subscripts identify
the atomic levels involved in a transition, while the superscripts 
represent the frequencies of the corresponding emitted ($+$) and 
absorbed ($-$) photons. These profiles originate in our theory from 
the formal product 
of a first-order Feynman diagram with a second-order diagram. The two
diagrams are distinguished by the comma separators in both the subscript
and superscript lists. The time ordering of these Feynman diagrams is 
from right to left, in the profile's notation. Thus, in 
Equation~(\ref{eq:Psi}), we read the contribution 
of a first-order process involving the transition $b\to a$ with exchange
of a photon $\bm{h}$, and at the same time the contribution of a second-order
process involving the transition $e\to d\to c$ with exchange of two
photons, $\bm{l}$ and $\bm{k}$.

A pictorial representation of the
Feynman diagrams for atom-photon interactions up to second order is given 
in Figure~\ref{fig:diagrams}. In all diagrams, time is flowing upward. 
The first-order diagrams (top row) account for the processes of 
single-photon absorption and emission, while the second-order diagrams 
(bottom row) describe the scattering of photons (first two diagrams), 
as well as two-photon absorption and emission (last two diagrams). The 
central diagram in the bottom row represents a typical atomic self-energy 
insertion, corresponding to the lowest-order radiative correction to the 
energy of the atomic state. This diagram is accounted for in the theory 
by ``dressing'' the atomic propagator, which results in the atomic levels 
acquiring a finite lifetime because of the possibility of spontaneous 
de-excitation. 

The process of two-photon absorption is the simplest example of 
non-linear optical effect, and can typically be neglected for highly 
diluted radiation fields, such as in the case of solar applications.
Diagrams of that type give rise to radiation observables that are
related to \emph{second-order coherence} \cite[e.g.,][]{Lo83,MW95}, 
and which we explicitly neglect in our treatment, although the impact 
of this choice for the inner consistency of the formalism is still to be 
fully assessed.

We will also ignore the two-photon emission diagram of
Fig.~\ref{fig:diagrams}, as this does not contribute significantly to 
the scattering of radiation in a two-term atom model. In fact, because 
of energy conservation, the initial and final states in that diagram, 
$p$ and $q$, must belong respectively to the upper and lower terms of 
the model atom, and so the intermediate transition through the 
virtual state $r$ is constrained by the set of selection rules
that apply to the $p$--$q$ transition. This typically results in a very
low rate for the two-photon emission process. An instructive example 
is that of the Lyman-$\alpha$ transition of hydrogen, between the atomic 
terms of principal quantum numbers $n=2$ and $n=1$. The Lyman-$\alpha$
emission is dominated by the electric-dipole transition from
$2\,P_{1/2,3/2}$ to $1\,S_{1/2}$, with a lifetime 
$\approx 1.6\times 10^{-9}$\,s. The transition between $2\,S_{1/2}$ and
$1\,S_{1/2}$ is forbidden to the lowest order of approximation, but it 
can occur via two-photon emission, corresponding to the last diagram 
of Fig.~\ref{fig:diagrams}, through a virtual state $r$ belonging to 
a $n\,P_J$ term. 
Such transition has indeed been observed in the laboratory, and it is 
responsible for a measurable lifting of the metastability of the 
$1\,S_{1/2}$ state, with a lifetime $\approx 0.14$\,s \citep{BS57}. 
However, its contribution compared to the dominant term of 
Lyman-$\alpha$ is completely negligible for any practical application 
to the polarized line diagnostics of astrophysical plasmas.

In the general case of a multi-term atom, there are additional terms that 
must be considered, which originate from different combinations of 
Feynman diagrams (involving up to third-order processes, to the degree
of approximation of the theory). \emph{For the two-term 
atom, and neglecting stimulation effects, the profiles of 
Equation~(\ref{eq:Psi}) are the only ones bringing a contribution to the 
scattering of polarized radiation.}% \textbf{\uppercase{is this true?}}

After a rather involved algebraic manipulation, the sum of complex 
profiles in Equation~(\ref{eq:2emiss}) is shown to be proportional to the 
following redistribution function,
\begin{eqnarray}	\label{eq:red.gen}
&&\kern -1cm
{\cal R}(\Omega_u,\Omega_{u'};
	\Omega_l,\Omega_{l'},\Omega_{l''};
	\omega_k,\omega_{k'})
	\equiv  
	(\epsilon_{uu'}+{\rm i}\omega_{uu'})
	\left(
	\Psi_{u'l',l''ul}^{-k,+k'-k} + \bar\Psi_{ul,l''u'l'}^{-k,+k'-k}
	\right)
\nonumber \\ 
&=&
\frac{2\epsilon_{l''}(\epsilon_{ll'}+{\rm i}\omega_{ll'})}{%
	(\omega_k-\omega_{ul'}+{\rm i}\epsilon_{ul'})
       (\omega_k-\omega_{u'l}-{\rm i}\epsilon_{u'l})
	(\omega_{k'}-\omega_{ul''}+{\rm i}\epsilon_{ul''})
	(\omega_{k'}-\omega_{u'l''}-{\rm i}\epsilon_{u'l''})}
\nonumber \\
&+&
\frac{2\epsilon_{l''}
	(\epsilon_{uu'}+{\rm i}\omega_{uu'})}{%
	(\omega_k-\omega_{ul'}+{\rm i}\epsilon_{ul'})
	(\omega_k-\omega_{u'l}-{\rm i}\epsilon_{u'l})
	(\omega_k-\omega_{k'}+\omega_{l'l''}+{\rm i}\epsilon_{l'l''})
	(\omega_k-\omega_{k'}+\omega_{ll''}-{\rm i}\epsilon_{ll''})}
\nonumber \\
&+&
\frac{(\epsilon_{ll'}+{\rm i}\omega_{ll'})
	(\epsilon_{uu'}+{\rm i}\omega_{uu'})}{%
	(\omega_{k'}-\omega_{ul''}+{\rm i}\epsilon_{ul''})
	(\omega_{k'}-\omega_{u'l''}-{\rm i}\epsilon_{u'l''})
	(\omega_k-\omega_{k'}+\omega_{l'l''}+{\rm i}\epsilon_{l'l''})
	(\omega_k-\omega_{k'}+\omega_{ll''}-{\rm i}\epsilon_{ll''})}
\nonumber \\
&+&
\frac{2\epsilon_{l''}
	(\epsilon_{ll'}+{\rm i}\omega_{ll'})
	(\epsilon_{uu'}+{\rm i}\omega_{uu'})}{%
	(\omega_k-\omega_{ul'}+{\rm i}\epsilon_{ul'})
	(\omega_k-\omega_{u'l}-{\rm i}\epsilon_{u'l})
	(\omega_{k'}-\omega_{ul''}+{\rm i}\epsilon_{ul''})
	(\omega_{k'}-\omega_{u'l''}-{\rm i}\epsilon_{u'l''})}
\nonumber \\
&&\kern 1cm \times
\frac{2\epsilon_{l''}+\epsilon_{ll'}+\epsilon_{uu'}
      +{\rm i}(\omega_{ll'}+\omega_{uu'})}{%
	(\omega_k-\omega_{k'}+\omega_{l'l''}+{\rm i}\epsilon_{l'l''})
	(\omega_k-\omega_{k'}+\omega_{ll''}-{\rm i}\epsilon_{ll''})}\;,
\end{eqnarray}
where we introduced the sum of the level widths, 
$\epsilon_{ab}=\epsilon_a+\epsilon_b$, and in the argument list of the
function ${\cal R}$ we adopted the shorthand notation 
$\Omega_a=\omega_a-{\rm i}\epsilon_a$, to indicate the complex energy of 
the level $a$, including its width.

Equation~(\ref{eq:red.gen}) is one of the preliminary results of our
theory of partial redistribution with polarization. Another result 
suggests that, \emph{in the absence of collisions 
(i.e., at zero temperature), the process that populates the upper term 
by radiative absorption is inhibited in the limit of infinitely sharp 
lower levels.} In this limit, the first-order contribution to the 
emissivity in Equation~(\ref{eq:RT}) vanishes, and all radiation processes 
contributing to the emergent radiation depend exclusively on the population
distribution and coherence within the lower term.
The self-consistency of this last result has not yet been fully 
demonstrated, and for this reason we cannot provide here a rigorous proof 
of the phenomenon of inhibition of one-photon absorption in the limit of 
infinitely sharp lower levels. However, such picture is supported also by the following 
intuitive physical argument. Since the ideal limit of sharp lower levels 
is equivalent to that of infinite lifetime of the same 
levels, any measurement aimed at determining the population distribution 
of the atomic system would always find the system in the lower state. 
In other words, any radiative transition involving the upper levels can 
only be virtual, without actually populating the upper state (in the
sense specified above). Thus the corresponding contribution to the emitted 
radiation must be fully accounted for by the coherent-scattering 
emissivity given in Equation~(\ref{eq:2emiss}), in which only the
density matrix of the lower state explicitly appears.

Equation (\ref{eq:red.gen}) represents the most general form 
of redistribution function for a two-term atom with polarized levels
that are partially degenerate or non degenerate. It includes, as 
particular cases, the redistribution
functions for the two- and three-level atoms that have formerly been
described in the literature (see Section~\ref{sec:special}).
While it is possible to cast 
${\cal R}(\Omega_u,\Omega_{u'}; \Omega_l,\Omega_{l'},\Omega_{l''};
	\omega_k,\omega_{k'})$ in various equivalent forms (see, e.g., 
Equation~(\ref{eq:red.gen.alt}) below) -- all of them descending directly 
from Equation~(\ref{eq:Psi}) and the definition in Equation~(\ref{eq:red.gen}) -- 
the particular form that we adopted for Equation~(\ref{eq:red.gen}) is 
best suited for re-deriving those former results in the partial 
redistribution literature, as we are now going to show. 

%\begin{figure}[h!]
%\includegraphics[width=\hsize]{OSC_F.eps}
%\end{figure}

\section{Special forms of the redistribution function} \label{sec:special}
We consider in this section a few particular cases of
Equation~(\ref{eq:red.gen}), some of which correspond to well-known results 
from the past literature on partial redistribution.
The simplest case is that of a transition between two atomic levels with 
fully degenerate sublevels (i.e., for zero magnetic field), so that 
$\Omega_u=\Omega_{u'}$ and 
$\Omega_l=\Omega_{l'}=\Omega_{l''}$. From Equation~(\ref{eq:red.gen}), 
we immediately find
\begin{eqnarray}	\label{eq:red.0}
{\cal R}_0(\Omega_u; \Omega_l; \omega_k,\omega_{k'}) 
&=&
\frac{4\epsilon_l^2}{%
	\left[(\omega_k-\omega_{ul})^2+\epsilon_{ul}^2\right]\!\!
	\left[(\omega_{k'}-\omega_{ul})^2+\epsilon_{ul}^2\right]}
\nonumber \\
&&\kern -1in
\mathop{+}
\frac{4\epsilon_l \epsilon_u}{%
	\left[(\omega_k-\omega_{ul})^2+\epsilon_{ul}^2\right]\!\!
	\left[(\omega_k-\omega_{k'})^2+4\epsilon_l^2\right]} 
%\nonumber \\
%&&\mathop{+}
	+\frac{4\epsilon_l \epsilon_u}{%
	\left[(\omega_{k'}-\omega_{ul})^2+\epsilon_{ul}^2\right]\!\!
	\left[(\omega_k-\omega_{k'})^2+4\epsilon_l^2\right]}
\nonumber \\
&&\kern -1in
\mathop{+}
\frac{16\epsilon_l^2\epsilon_u(2\epsilon_l+\epsilon_u)}{%
	\left[(\omega_k-\omega_{ul})^2+\epsilon_{ul}^2\right]\!\!
	\left[(\omega_{k'}-\omega_{ul})^2+\epsilon_{ul}^2\right]\!\!
	\left[(\omega_k-\omega_{k'})^2+4\epsilon_l^2\right]}\;,
\end{eqnarray}
which is identical to the redistribution function derived by 
\citet[][cf.~their Equation~(VIII.3.33); see also \citealt{Om72}, 
and \citealt{Mi78}]{WS53} using a metalevel model of the atomic system.

The case of a three-level ``$\Lambda$-type'' atomic system \cite[see,
e.g.,][Figure~2]{LT71}
with fully degenerate sublevels, is obtained by letting 
$\Omega_u=\Omega_{u'}$ and
$\Omega_l=\Omega_{l'}\ne\Omega_{l''}$ in Equation~(\ref{eq:red.gen}), 
which gives at once
\begin{eqnarray}	\label{eq:red.1}
{\cal R}_1(\Omega_u; \Omega_l,\Omega_{l''}; 
	\omega_k,\omega_{k'}) 
&=&
\frac{4\epsilon_l\epsilon_{l''}}{%
      \left[(\omega_k-\omega_{ul})^2+\epsilon_{ul}^2\right]\!\!
	\left[(\omega_{k'}-\omega_{ul''})^2+\epsilon_{ul''}^2\right]}
\nonumber \\
&&\kern -4cm
\mathop{+}
\frac{4\epsilon_{l''}\epsilon_u}{%
	\left[(\omega_k-\omega_{ul})^2+\epsilon_{ul}^2\right]\!\!
	\left[(\omega_k-\omega_{k'}+\omega_{ll''})^2+\epsilon_{ll''}^2\right]}
%\nonumber \\
%&&\kern -2cm
%\mathop{+}
+\frac{4\epsilon_l\epsilon_u}{%
	\left[(\omega_{k'}-\omega_{ul''})^2+\epsilon_{ul''}^2\right]\!\!
	\left[(\omega_k-\omega_{k'}+\omega_{ll''})^2+\epsilon_{ll''}^2\right]}
\nonumber \\
&&\kern -4cm
\mathop{+}
\frac{16\epsilon_l\epsilon_{l''}\epsilon_u(\epsilon_{ll''}+\epsilon_u)}{%
	\left[(\omega_k-\omega_{ul})^2+\epsilon_{ul}^2\right]\!\!
	\left[(\omega_{k'}-\omega_{ul''})^2+\epsilon_{ul''}^2\right]\!\!
	\left[(\omega_k-\omega_{k'}+\omega_{ll''})^2+\epsilon_{ll''}^2\right]}\;.
\end{eqnarray}
This problem, which describes, for example, the Raman scattering in 
subordinate lines, has also been previously considered in the 
literature, 
e.g., by \citet[][Equation~(3.2); see also \citealt{WS53}, and 
\citealt{We33}]{Hu82}.
%
%More generally, Equations~(\ref{eq:red.0}) and (\ref{eq:red.1}) also describe 
%the cases of Rayleigh and Raman scattering, respectively, in a multi-level
%atom with degenerate and unpolarized levels \textbf{(??? non so bene
%cosa volessi dire qui ???)}.

Of more general interest is the redistribution function for a two-term 
atom with non-degenerate levels, and completely relaxed atomic 
coherence in the lower term (\textit{non-coherent lower term}; 
n.c.l.t.). Such a model can 
describe the formation of many resonance lines observed in the quiet-Sun 
atmosphere, in a regime of magnetic strengths such that the Hanle effect 
of the lower levels is saturated, but not so large as to induce 
level-crossing interference in the lower term. For example, the 
formation
of the \ion{Na}{1} D$_1$-D$_2$ doublet at $\lambda 589$\,nm falls into 
this category, for $0.1\,\mathrm{G}\lesssim B\lesssim 300\,\textrm{G}$
\citep{TB02,Ca02}. In order to derive this form of the redistribution
function, we observe that the diagonality condition 
$\rho_{ll'}=\delta_{ll'}\rho_{ll}$ for the case of n.c.l.t.\ implies 
$\Omega_l=\Omega_{l'}$ in Equation~(\ref{eq:red.gen}). We thus obtain
\begin{eqnarray}	\label{eq:red.nclt}
&&\kern -1cm
{\cal R}(\Omega_u,\Omega_{u'};
	\Omega_l,\Omega_{l''};
	\omega_k,\omega_{k'})_{\hbox{\footnotesize n.c.l.t.}} 
\nonumber \\ 
&=&
\frac{4\epsilon_l\epsilon_{l''}}{%
	(\omega_k-\omega_{ul}+{\rm i}\epsilon_{ul})
       (\omega_k-\omega_{u'l}-{\rm i}\epsilon_{u'l})
	(\omega_{k'}-\omega_{ul''}+{\rm i}\epsilon_{ul''})
	(\omega_{k'}-\omega_{u'l''}-{\rm i}\epsilon_{u'l''})}
\nonumber \\
&+&
\frac{2\epsilon_{l''}
	(\epsilon_{uu'}+{\rm i}\omega_{uu'})}{%
	(\omega_k-\omega_{ul}+{\rm i}\epsilon_{ul})
	(\omega_k-\omega_{u'l}-{\rm i}\epsilon_{u'l})
	\left[(\omega_k-\omega_{k'}+\omega_{ll''})^2+\epsilon_{ll''}^2\right]}
\nonumber \\
&+&
\frac{2\epsilon_l
	(\epsilon_{uu'}+{\rm i}\omega_{uu'})}{%
	(\omega_{k'}-\omega_{ul''}+{\rm i}\epsilon_{ul''})
	(\omega_{k'}-\omega_{u'l''}-{\rm i}\epsilon_{u'l''})
	\left[(\omega_k-\omega_{k'}+\omega_{ll''})^2+\epsilon_{ll''}^2\right]}
\nonumber \\
&+&
\frac{4\epsilon_l\epsilon_{l''}
	(\epsilon_{uu'}+{\rm i}\omega_{uu'})}{%
	(\omega_k-\omega_{ul}+{\rm i}\epsilon_{ul})
	(\omega_k-\omega_{u'l}-{\rm i}\epsilon_{u'l})
	(\omega_{k'}-\omega_{ul''}+{\rm i}\epsilon_{ul''})
	(\omega_{k'}-\omega_{u'l''}-{\rm i}\epsilon_{u'l''})}
\nonumber \\
&&\kern 1cm \times
\frac{2\epsilon_{l''}+2\epsilon_l+\epsilon_{uu'}
	+{\rm i}\omega_{uu'}}{%
	(\omega_k-\omega_{k'}+\omega_{ll''})^2+\epsilon_{ll''}^2}\;.
\end{eqnarray}
Often the initial and final levels of the scattering process, $l$ and 
$l''$, represent metastable states, hence characterized by very small 
radiative level widths. In this limit of \emph{sharp lower levels} (s.l.l.),
$\epsilon_l,\epsilon_{l''}\to 0$, the first and fourth contributions 
in Equation~(\ref{eq:red.nclt}) vanish, and the redistribution function 
tends to the well-known result \cite[e.g.,][]{La97}
\begin{eqnarray}	\label{eq:red.sll}
{\cal R}_2(\Omega_u,\Omega_{u'};
	\Omega_l,\Omega_{l''};
	\omega_k,\omega_{k'})
&=& \frac{2\pi\,
	(\epsilon_{uu'}+{\rm i}\omega_{uu'})\,
	\delta(\omega_k-\omega_{k'}+\omega_{ll''})}{%
	(\omega_k-\omega_{ul}+{\rm i}\epsilon_{u})
	(\omega_k-\omega_{u'l}-{\rm i}\epsilon_{u'})}\;,
\end{eqnarray}
where we also considered that
$\omega_{k'}+\omega_{l''}=\omega_k+\omega_l$
(because of the Dirac-$\delta$ function) in order to combine the two 
non-vanishing contributions of Equation~(\ref{eq:red.nclt}). 
%Despite the apparent simplicity of its form, Equation~(\ref{eq:red.sll}) 
%is already sufficient to describe partial redistribution effects in 
%polarized scattering for all models of the two-term atom, even in 
%the presence of a magnetic field, at the condition that the magnetic 
%strength is such that the Hanle effect for the lower term is saturated, 
%but not large enough to cause level crossing (in case that the lower 
%term has more than one level).

A more general case is that of infinitely sharp lower levels with 
the possibility of atomic coherence in the lower term. 
The derivation of the corresponding redistribution function from 
Equation~(\ref{eq:red.gen}) is not trivial. It is however easily 
accomplished if we first recast the general redistribution function 
of Equation~(\ref{eq:red.gen}) in the alternative but equivalent form
\begin{eqnarray}	\label{eq:red.gen.alt}
&&\kern -1cm
{\cal R}(\Omega_u,\Omega_{u'};
	\Omega_l,\Omega_{l'},\Omega_{l''};
	\omega_k,\omega_{k'}) \nonumber \\ 
&=&(\epsilon_{uu'}+{\rm i}\omega_{uu'})\biggl[
	\frac{{\rm i}}{%
	(\omega_k-\omega_{ul}+{\rm i}\epsilon_{u}-{\rm i}\epsilon_{l})
	(\omega_k-\omega_{u'l}-{\rm i}\epsilon_{u'l})
	(\omega_{k'}-\omega_k-\omega_{ll''}+{\rm i}\epsilon_{ll''})}
	\nonumber \\
&&\hphantom{(\epsilon_{uu'}+{\rm i}\omega_{uu'})\biggl[}-
	\frac{{\rm i}}{%
	(\omega_k-\omega_{ul'}+{\rm i}\epsilon_{ul'})
	(\omega_k-\omega_{u'l'}-{\rm i}\epsilon_{u'}+{\rm i}\epsilon_{l'})
	(\omega_{k'}-\omega_k-\omega_{l'l''}-{\rm i}\epsilon_{l'l''})}
	\biggr] \nonumber \\
%\noalign{\eject}
&+&(\epsilon_{ll'}+{\rm i}\omega_{ll'})\biggl[
	\frac{{\rm i}}{%
	(\omega_k-\omega_{ul}+{\rm i}\epsilon_{u}-{\rm i}\epsilon_{l})
	(\omega_k-\omega_{ul'}+{\rm i}\epsilon_{ul'})
	(\omega_{k'}-\omega_{ul''}+{\rm i}\epsilon_{ul''})}
	\nonumber \\
&&\hphantom{(\epsilon_{ll'}+{\rm i}\omega_{ll'})\biggl[}-
	\frac{{\rm i}}{%
	(\omega_k-\omega_{u'l}-{\rm i}\epsilon_{u'l})
	(\omega_k-\omega_{u'l'}-{\rm i}\epsilon_{u'}+{\rm i}\epsilon_{l'})
	(\omega_{k'}-\omega_{u'l''}-{\rm i}\epsilon_{u'l''})}
	\biggr]\;.
\end{eqnarray}
We also define the generalized function\footnote{This definition 
differs from the $\zeta(x)$ function introduced by \cite{He54} for
the presence of ${\rm i}$ at the numerator.}
\begin{equation} \label{eq:zeta}
\zeta(x)\equiv\lim_{\sigma\to 0}\,\frac{{\rm i}}{x+{\rm i}\,\sigma}
	=\pi\,\delta(x)+{\rm i}\;{\rm Pv}\,\frac{1}{x}\;.
\end{equation}
Then, in the limit of $\epsilon_l,\epsilon_{l'},\epsilon_{l''}\to 0$, 
the redistribution function for sharp lower levels
follows at once from Equation~(\ref{eq:red.gen.alt}):
\begin{eqnarray} \label{eq:red.sharp}
&&\kern -1cm
{\cal R}(\Omega_u,\Omega_{u'};
	\Omega_l,\Omega_{l'},\Omega_{l''};
	\omega_k,\omega_{k'})_{\rm s.l.l.} \nonumber \\
\noalign{\vskip 3pt}
&=&(\epsilon_{uu'}+{\rm i}\omega_{uu'})
	\biggl[
	\frac{\zeta(\omega_k-\omega_{k'}+\omega_{ll''})}{%
		(\omega_k-\omega_{ul}+{\rm i}\epsilon_{u})
		(\omega_k-\omega_{u'l}-{\rm i}\epsilon_{u'})}
	+\frac{\zeta^*(\omega_k-\omega_{k'}+\omega_{l'l''})}{%
		(\omega_k-\omega_{ul'}+{\rm i}\epsilon_{u})
		(\omega_k-\omega_{u'l'}-{\rm i}\epsilon_{u'})}
	\biggr] \nonumber \\
\noalign{\vskip 6pt}
&-&\omega_{ll'} \biggl[
	\frac{1}{%
		(\omega_k-\omega_{ul}+{\rm i}\epsilon_{u})
		(\omega_k-\omega_{ul'}+{\rm i}\epsilon_{u})
		(\omega_{k'}-\omega_{ul''}+{\rm i}\epsilon_{u})}
	\nonumber \\
&&\hphantom{\omega_{ll'} \biggl[} -
	\frac{1}{%
		(\omega_k-\omega_{u'l}-{\rm i}\epsilon_{u'})
		(\omega_k-\omega_{u'l'}-{\rm i}\epsilon_{u'})
		(\omega_{k'}-\omega_{u'l''}-{\rm i}\epsilon_{u'})}
	\biggr]\;.
\end{eqnarray}
We note that Equation~(\ref{eq:red.sharp}) properly tends to the
redistribution function ${\cal R}_2$ of Equation~(\ref{eq:red.sll}) when 
$\omega_{ll'}\to 0$, since there is no contribution from the second square
bracket in this case, and $\zeta(x)+\zeta^*(x)=2\pi\,\delta(x)$.

We conclude this section with the evaluation of the integral norm
of the general redistribution function, Equation~(\ref{eq:red.gen}) or
(\ref{eq:red.gen.alt}). By contour 
integration,\footnote{\label{note:domainextend}%
Because $\omega_{ul}\gg\epsilon_{ul}$, the
frequency domain of integration $[0,+\infty)$ can always be extended 
backward to $-\infty$ with no appreciable modification of the value of 
the integral.} we find
\begin{eqnarray} \label{eq:integral2}
\int_{-\infty}^\infty d\omega_{k'}\;
	{\cal R}(\Omega_u,\Omega_{u'};
	\Omega_l,\Omega_{l'},\Omega_{l''};
	\omega_k,\omega_{k'})
%	\left(
%	\Psi_{u'l',l''ul}^{-k,+k'-k} + 
%	\bar\Psi_{ul,l''u'l'}^{-k,+k'-k}
%	\right)
&=& 2\pi\,
	\frac{\epsilon_{uu'}+\epsilon_{ll'}
	+{\rm i}(\omega_{uu'}+\omega_{ll'})}{%
	(\omega_k-\omega_{ul'}+{\rm i}\epsilon_{ul'})
      (\omega_k-\omega_{u'l}-{\rm i}\epsilon_{u'l})
	}\;, \\
\int_{-\infty}^\infty d\omega_{k}\;
	{\cal R}(\Omega_u,\Omega_{u'};
	\Omega_l,\Omega_{l'},\Omega_{l''};
	\omega_k,\omega_{k'})
%	\left(
%	\Psi_{u'l',l''ul}^{-k,+k'-k} + 
%	\bar\Psi_{ul,l''u'l'}^{-k,+k'-k}
%	\right)
&=& 2\pi\,
	\frac{\epsilon_{uu'}+2\epsilon_{l''}
	+{\rm i}\omega_{uu'}}{%
	(\omega_{k'}-\omega_{ul''}+{\rm i}\epsilon_{ul''})
      (\omega_{k'}-\omega_{u'l''}-{\rm i}\epsilon_{u'l''})
	}\;,
\end{eqnarray}
and from either of these two expressions it follows immediately that
\begin{equation} \label{eq:red.norm}
\int_{-\infty}^\infty d\omega_k
\int_{-\infty}^\infty d\omega_{k'}\,
	{\cal R}(\Omega_u,\Omega_{u'};
	\Omega_l,\Omega_{l'},\Omega_{l''};
	\omega_k,\omega_{k'})
= 4\pi^2\;.
\end{equation}

The results expressed by Equations~(\ref{eq:integral2})--(\ref{eq:red.norm}) 
differ from those derived in the theory of \cite{Bo97a,Bo97b}, where the 
line shape function in the radiative transfer equation contributed by 
higher orders is shown to vanish, when integrated over the frequency of 
either the incoming or the outgoing photon. On the other hand, in Bommier's 
formalism, the contribution of the first-order emissivity is always present, 
even when the lifetime of the lower term is assumed to be infinite and thus 
no upper-term excitation should be expected, based on the physical argument 
presented at the end of Section~\ref{sec:theory}. Thus, in the work of
\cite{Bo97a}, the redistribution function always results from
the combination of both first- and second-order emissivity terms, even
in the limit of infinitely sharp lower levels. 
Accordingly,
redistribution effects are accounted for through a modified emission
coefficient where the distinction between true absorption and 
scattering is lost \cite[see also][]{Co82}.

Despite this difference
between our respective formalisms, 
in the collisionless regime and for an unpolarized lower
term, 
we verified that the redistribution function 
${\cal R}(\Omega_u,\Omega_{u'};
	\Omega_l,\Omega_{l''};
	\omega_k,\omega_{k'})_{\hbox{\footnotesize n.c.l.t.}}$ of
Equation~(\ref{eq:red.nclt}) coincides with that derived by 
\citeauthor{Bo97b} (\citeyear{Bo97b}, Equations~(43)--(45); for this 
demonstration, all level widths in those formulas that correspond 
to our $\epsilon_u$ must be replaced by $\epsilon_{ul}$).

The general redistribution function, Equation~(\ref{eq:red.gen}) 
or (\ref{eq:red.gen.alt}), with all the particular cases considered 
in this section, is expressed in the reference frame of the atom. In 
realistic applications to the plasma diagnostics of astrophysical
objects, we must generalize this function to the laboratory frame, so 
to include the effects of Doppler broadening through the convolution of 
the atomic velocity distribution (typical, a Maxwellian). In order not
to burden the presentation of the following discussion, we provide the 
details of such derivation in the Appendix.

\section{Interlude: the problem of radiative lifetimes}
\label{sec:lifetimes}

One critical question, in the application of Equations~(\ref{eq:2emiss}) and
(\ref{eq:red.gen}) (or (\ref{eq:red.gen.alt})) to the study of 
partial redistribution effects in 
the presence of external fields, is the proper definition of the level 
widths, $\epsilon_a$, i.e., of the lifetimes of atomic states that are 
in general energetically perturbed by the presence of external fields. 
If we restrict our considerations to the case of a two-term atom
with sharp lower levels, the problem of such definition concerns only the 
excited states $u$ and $u'$ (cf.~Equation~(\ref{eq:red.sharp})). To
the lowest order of approximation, their widths are determined exclusively 
by the external fields and by the process of spontaneous de-excitation 
(in particular, we are neglecting in this treatment the effects of stimulated 
emission). Within such an approximation, the complex frequencies $\Omega_u$ 
correspond to the eigenvalues of the non-Hermitian operator 
\cite[e.g.,][]{CT92} 
\begin{equation} \label{eq:non-Hermitean}
K_u=\mathscr{P}_u H_{\rm A}\mathscr{P}_u-{\rm i}\hbar\mathit{\Gamma}_u\;,
\end{equation}
where $H_{\rm A}$ is the Hamiltonian of the atomic system in the
presence of the external fields, $\mathscr{P}_u$ is the projection
operator over the subspace of the states $u$ of that Hamiltonian, 
and finally
\begin{equation} \label{eq:damping}
\mathit{\Gamma}_u=\frac{2}{3}\frac{e_0^2}{\hbar c^3}\,\omega_{ul}^3
	\sum_q \mathscr{P}_u\,r_q\,\mathscr{P}_l\,r_q^\dagger\,
      \mathscr{P}_u\;,
\end{equation}
where evidently $\mathscr{P}_l$ is the projection operator over the
subspace of the lower terms, which is radiatively connected to the 
subspace of the excited states $u$.

We note how in Equation~(\ref{eq:damping}) we have assumed that the energy 
separation between the levels of the upper and lower terms can be 
approximated by a single value $\omega_{ul}$. This is an adequate 
approximation in most cases of interest for magnetic studies 
of the solar atmosphere, where the field strengths at play, and the fine
structure of the atomic terms, are such that the associated energy
separations of the levels are much smaller than $\omega_{ul}$. 
Correspondingly, level mixing and 
quantum interference are also triggered between relatively close
levels, such that the energy span of the atomic terms (intended here 
as the sets of the interfering atomic levels) is much smaller than 
their average separation $\omega_{ul}$. Under this approximation, and 
in the presence of a magnetic field, it is possible to show that the 
\textit{damping matrix} $\mathit{\Gamma}_u$ is diagonal on the basis of the 
eigenvectors of $H_{\rm A}$ \cite[e.g.,][]{RC61}. In this case, $K_u$ 
is also diagonal on the same basis.
Thus, \emph{the level widths for radiative de-excitation correspond to 
the diagonal elements of the damping matrix} $\mathit{\Gamma}_u$.

The use of Equation~(\ref{eq:damping}) for determining the line shape of
the scattered radiation is adequate when the frequency of the emitted 
photon is near the atomic resonance. On the other hand, away from the 
resonance condition (e.g., in the case of Rayleigh scattering in the 
far wings of a line), it is questionable whether the lifetime of the 
excited state should contain the resonance frequency $\omega_{ul}$. 
The treatment of the natural line shape given by \cite{Po64} suggests 
in fact that one should replace $\omega_{k'}$ for $\omega_{ul}$ in the 
expression of $\mathit{\Gamma}_u$ (cf., in particular, Equations~(8.53) and (8.61) 
in \citealt{Po64}). Of course, in the proximity of the atomic resonance 
(i.e., on the energy shell), one retrieves the usual form of the 
damping matrix as given by Equation~(\ref{eq:damping}).

The problem of the lifetime of the lower term is significantly more
involved. First of all, it is important to remark that, because of the
limitations of our current theory of radiation scattering in the 
absence of collisions, no new contributions can arise that are able 
to ``dress'' a metastable lower term, and determine thus its radiative 
lifetime for transitions towards the upper level. Nonetheless, the 
energy widths $\epsilon_l$ of the lower term 
properly appear in the general form of the redistribution function, 
since those widths are associated with the complex frequencies 
$\Omega_l$. In the case of subordinate lines, for example, we can
expect that those widths will have an important contribution from 
spontaneous de-excitation, which is again determined through 
Equation~(\ref{eq:damping}).
In the case of a metastable state, instead, the radiative level widths can 
only be specified \emph{ad hoc}, since they are not provided
self-consistently by the theory. In real cases, such as for a collisional 
plasma at a temperature $T>0$, we can expect that the level widths of 
a metastable term will predominantly be determined by collisional processes. 
In the following, we will then assume that the radiative broadening of
metastable levels can be neglected, and therefore that the radiative
lifetime of such levels is practically infinite.

%\newpage

\section{The two-term atom with hyperfine structure}
\label{sec:application}

Standard manipulations \cite[e.g.,][]{LL04} allow us to express 
Equation~(\ref{eq:RT}) by means of the irreducible spherical tensor 
formalism, and the emerging polarized radiation $I_{\lambda'\mu'}(\bm{k'})$
in terms of the corresponding Stokes vector,
\begin{equation} \label{eq:RT.1}
\frac{1}{c}\,\frac{d}{dt}\,S_i(\omega_{k'},\bm{\hat{k}'}) =
	-\sum_j \kappa_{ij}(\omega_{k'},\bm{\hat{k}'})\,
		S_j(\omega_{k'},\bm{\hat{k}'})
	+\varepsilon\apx{1}_i(\omega_{k'},\bm{\hat{k}'})
	+\varepsilon\apx{2}_i(\omega_{k'},\bm{\hat{k}'})\;,\qquad (i=0,1,2,3)
\end{equation}
where
\begin{eqnarray} \label{eq:2emiss.1}
\varepsilon\apx{2}_i(\omega_{k'},\bm{\hat{k}'})
&\equiv&\frac{4}{3}\frac{e_0^4}{\hbar^2 c^4}\,{\cal N}\omega_{k'}^4
	\sum_{ll'}\rho_{ll'}\sum_{uu'l''}
	\sum_{qq'}\sum_{pp'}(-1)^{q'+p'}\,
	(r_q)_{ul}(r_{q'})^\ast_{u'l'}
	(r_p)_{u'l''}(r_{p'})^\ast_{ul''} \\
&&\times
	\sum_{KQ}\sum_{K'Q'}\sqrt{(2K+1)(2K'+1)}\,
	\thrj{1}{1}{K}{-q}{q'}{-Q}
	\thrj{1}{1}{K'}{-p}{p'}{-Q'}\,
	T^{K'}_{Q'}(i,\bm{\hat{k}'}) \nonumber \\
&&\times \int_0^\infty d\omega_k
	\left(
	\Psi_{u'l',l''ul}^{-k,+k'-k} + \bar\Psi_{ul,l''u'l'}^{-k,+k'-k}
	\right)
%	{\cal R}(\Omega_u,\Omega_{u'};
%	\Omega_l,\Omega_{l'},\Omega_{l''};
%	\omega_k,\omega_{k'})\,
	J^K_Q(\omega_k)\;.\qquad (i=0,1,2,3) \nonumber
\end{eqnarray}
%
%where the 1st-order terms include the ordinary contributions due to 
%the absorption and emission %(possibly corrected for stimulated effects)
%of radiation along the propagation direction $\bm{\hat{k}'}$.  
%
The geometric tensors $T^K_Q(i,\bm{\hat{k}})$ were introduced by
\cite{La84}, and are conveniently tabulated by 
\citeauthor{Bo97b} (\citeyear{Bo97b}; see also \citealt{LL04}). 
The radiation tensors $J^K_Q(\omega_k)$ are defined in terms of 
the incident Stokes vector as follows,
\begin{equation} \label{eq:Jtens}
J^K_Q(\omega_k)=\oint \frac{d\bm{\hat{k}}}{4\pi}\,
      \sum_{j=0}^3 T^K_Q(j,\bm{\hat{k}})\,S_j(\omega_k,\bm{\hat{k}})\;.
\end{equation}

It is instructive to compare the form of Equation~(\ref{eq:RT.1}) with 
the usual expression of the radiative transfer equation for unpolarized 
light \cite[e.g.,][]{Mi78,Sh91} including the contribution from 
non-monochromatic scattering,
\begin{eqnarray} \label{eq:scalar.RT}
\frac{1}{c}\,\frac{d}{dt}\,I(\omega_{k'},\bm{\hat{k}'}) =
	&-& \bigl[\kappa^\mathrm{abs}(\omega_{k'}) +
		 \kappa^\mathrm{sca}(\omega_{k'})\bigr]
		I(\omega_{k'},\bm{\hat{k}'}) \nonumber \\
	&+&\varepsilon(\omega_{k'},\bm{\hat{k}'})
	+\oint \frac{d\bm{\hat{k}}}{4\pi} \int_0^\infty d\omega_k\;
	 \chi(\omega_{k'},\bm{\hat{k}'};\omega_k,\bm{\hat{k}})\,
	I(\omega_k,\bm{\hat{k}})\;.
\end{eqnarray}
We see that the last term with the double integral corresponds to 
$\varepsilon\apx{2}_i(\omega_{k'},\bm{\hat{k}'})$ of
Equation~(\ref{eq:2emiss.1}) for the case of unpolarized 
radiation, as it becomes evident when we
specialize Equation~(\ref{eq:Jtens}) to that case (i.e.,
$S_1=S_2=S_3=0$). In particular, for monochromatic
(e.g., Rayleigh) scattering, 
$\chi(\omega_{k'},\bm{\hat{k}'};\omega_k,\bm{\hat{k}})\equiv
\chi_{\omega_{k'}}(\bm{\hat{k}'},\bm{\hat{k}})\,\delta(\omega_k-\omega_{k'})$,
and Equation~(\ref{eq:scalar.RT}) becomes formally identical to Equation~(1.19) of 
\cite{Sh91}.

In the particular case of sharp lower levels, and in the
absence of collisions, the 1st-order emission term in the transfer 
equation vanishes. As we anticipated in the discussion following
Equation~(\ref{eq:red.gen}), this condition follows from the derivation 
of the statistical equilibrium equations according to the formalism on 
which the results presented in this paper are based. 
Thus, in the case of sharp lower levels, and in the absence of
collisions, the only 1st-order contribution to Equation~(\ref{eq:RT.1}) 
is represented by the absorption term.
Because there is no excitation of the upper levels (due to the infinite
lifetime of the lower levels), in this case the absorption term does not 
originate from true photon absorption, and corresponds instead to the atom's 
cross-section for the coherent scattering of radiation 
(see Sect.~\ref{sec:balance}). In contrast, when
the lifetime of the lower levels for transitions towards 
the upper term is finite, one must also account for the emissivity 
term of the first order in Equation~(\ref{eq:RT.1}), which is associated 
with the spontaneous de-excitation of the atom following the %(inelastic) 
excitation of the upper term.
%
%In that case, the absorption term must then 
%correspond to the \emph{total} cross-section for radiation scattering, 
%%accounting for both coherent and incoherent (i.e., inelastic, or from
%accounting for both coherent and incoherent (i.e., 
%true absorption) contributions. We note that this interpretation of 
%the absorption coefficient is in agreement with the results of the optical 
%theorem \citep[e.g.,][Sect.~3.3]{Ro65}.

%As we clarified earlier, this term does not originate from true photon
%absorption, with a corresponding excitation of the upper term, because 
%of the infinite lifetime of the lower levels. Instead it corresponds 
%to the atom's cross-section for coherent (in this case, purely elastic) 
%scattering of radiation. If the radiative lifetime of the lower levels for
%transitions towards the upper term were finite, the absorption 
%term would then correspond to the \emph{total} cross-section for radiation 
%scattering, accounting also for the incoherent (i.e., inelastic,
%or true absorption) contribution, in addition to the coherent
%contribution. 
%We must note that this interpretation of the absorption term is in 
%agreement with the results of the optical theorem \citep[e.g.,][]{Ro65}.
%%
%Of course, in the general case where the lifetime of the lower levels 
%for transitions towards the upper term (whether radiative, collisional, 
%or both) is finite, one must also include in Equation~(\ref{eq:RT.1}) 
%the usual emissivity term of the first order, which is associated with 
%the spontaneous de-excitation of the atom from the upper term down to 
%the lower term.

In this section we want to specialize Equation~(\ref{eq:RT.1}) to the case of a 
two-term atom with hyperfine structure, in the presence of a magnetic 
field. This is a general enough model to encompass most of the 
chromospheric lines of interest for magnetic diagnostics. The formalism is
similar to that of \cite{CM05}, so we extensively use the same notation of 
that paper.

If we indicate
with $\alpha_l$ and $\alpha_u$ the electronic configurations of the
lower and upper terms, respectively, and assume the direction of the
magnetic field as the quantization axis ($z$-axis), then the atomic 
states involved in Equation~(\ref{eq:RT.1}) are of the form
\begin{eqnarray*}
l&\equiv&\alpha_l I\mu_l M_l\;,\quad
l'\equiv\alpha_l I\mu_l' M_l'\;,\quad
l''\equiv\alpha_l I\mu_l'' M_l''\;,\\
u&\equiv&\alpha_u I\mu_u M_u\;,\quad
u'\equiv\alpha_u I\mu_u' M_u'\;,
\end{eqnarray*}
where $M$ is the projection of the total angular momentum $\bm{F}$ on
the $z$-axis, $I$ is the quantum number of the nuclear spin, while 
$\mu$ is the index of the atomic Hamiltonian eigenbasis spanning the 
subspace of all the quantum numbers $J$ and $F$ that are associated with 
a given value of $M$.  We then can write
\begin{displaymath}
\ket{\alpha I\mu M}=\sum_{JF} C^{JF}_\mu(\alpha I M)\,
	\ket{\alpha IJFM}\;,
\end{displaymath}
with the orthogonality conditions
\begin{equation} \label{eq:diag.prop}
\sum_{JF} C^{JF}_\mu(\alpha I M)\,C^{JF}_{\mu'}(\alpha I M)
	=\delta_{\mu\mu'}\;,\qquad
\sum_\mu C^{JF}_\mu(\alpha I M)\,C^{J'F'}_\mu(\alpha I M)
	=\delta_{JJ'}\delta_{FF'}\;.
\end{equation}

The density matrix element for the lower state, $\rho_{ll'}$, can be
written in terms of the irreducible spherical tensor components of the
statistical operator,
\begin{eqnarray}
\rho_{ll'}
&\equiv&\bra{\alpha_l I\mu_l M_l}\rho\ket{\alpha_l I \mu_l' M_l'} 
	\nonumber \\
\noalign{\vspace{6pt}}
&=&\sum_{J_l F_l}\sum_{J_l' F_l'}
	C^{J_l F_l}_{\mu_l}(\alpha_l I M_l)\,
	C^{J_l' F_l'}_{\mu_l'}(\alpha_l I M_l') \nonumber \\
\noalign{\vspace{-6pt}}
&&\hphantom{\sum_{J_l F_l}\sum_{J_l' F_l'}} \times
	\sum_{K_l Q_l} (-1)^{F_l-M_l}\sqrt{2K_l+1}\,
	\thrj{F_l}{F_l'}{K_l}{M_l}{-M_l'}{-Q_l}\,
	{}^{\alpha_l I}\!\rho^{K_l}_{Q_l}(J_l F_l,J_l' F_l')\;.
\end{eqnarray}
Finally, using the Wigner-Eckart theorem and its corollaries
\citep[e.g.,][]{BS93}, we 
derive the following expression of the dipole matrix element,
\begin{eqnarray}
(r_q)_{ul}
&\equiv&\bra{\alpha_u I\mu_u M_u}r_q\ket{\alpha_l I \mu_l M_l} 
	\nonumber \\
\noalign{\vspace{6pt}}
&=&\sum_{J_u F_u}\sum_{J_l F_l}
	C^{J_u F_u}_{\mu_u}(\alpha_u I M_u)\,
	C^{J_l F_l}_{\mu_l}(\alpha_l I M_l)\,
	\sqrt{(2J_u+1)(2F_u+1)(2F_l+1)}\;
	\redmat{\alpha_u J_u}{\bm{r}}{\alpha_l J_l} \nonumber \\
\noalign{\vspace{-6pt}}
&&\hphantom{\sum_{J_u F_u}\sum_{J_l F_l}} \times
	(-1)^{I+J_u-M_u}
	\thrj{F_u}{F_l}{1}{-M_u}{M_l}{q}
	\sixj{F_u}{F_l}{1}{J_l}{J_u}{I}\;.
\end{eqnarray}

To further proceed, we will assume that the two-term atom is 
adequately described within the $LS$-coupling scheme, so that
$\alpha\equiv\beta LS$, where $\beta$ identifies a particular $LS$
term of the atom. We then can write additionally
\begin{eqnarray}
\redmat{\alpha_u J_u}{\bm{r}}{\alpha_l J_l}
&\equiv&
	\redmat{\beta_u L_u S J_u}{\bm{r}}{\beta_l L_l S J_l}\;.
	\nonumber \\
&=&(-1)^{1+L_u+S+J_l} \sqrt{(2L_u+1)(2J_l+1)}\;
	\sixj{J_u}{J_l}{1}{L_l}{L_u}{S}\,
	\redmat{\beta_u L_u}{\bm{r}}{\beta_l L_l}\;.
\end{eqnarray}
The reduced matrix element in the last line can be expressed
in terms of the Einstein $B_{lu}$ coefficient for absorption from the
lower to the upper term of the atom,
\begin{equation}
B_{lu}=\frac{16\pi^3}{3}\frac{e_0^2}{\hbar^2 c}\,
	\frac{2L_u+1}{2L_l+1}\,
	|\redmat{\beta_u L_u}{\bm{r}}{\beta_l L_l}|^2\;,
\end{equation}
or alternatively through the Einstein coefficient for spontaneous emission
(cf.~Equation~(\ref{eq:damping})),
\begin{equation} \label{eq:Einstein A}
A_{ul}
=\frac{4}{3}\frac{e_0^2}{\hbar c^3}\,
	\omega_{ul}^3\,
	|\redmat{\beta_u L_u}{\bm{r}}{\beta_l L_l}|^2
\equiv\frac{4}{3}\frac{e_0^2}{\hbar c^3}\,
	\omega_{ul}^3
	\sum_{ql}|(r_q)_{ul}|^2
\;.
\end{equation}

It is relevant at this point to follow up on the discussion 
presented at the end of Section~\ref{sec:lifetimes}, with regard to the 
proper expression that must be adopted for the natural width of the 
excited state $u$, under excitation conditions that are away from the 
atomic resonance. 
Because of the conclusions of that discussion, we note that we cannot 
make the usual identification $A_{ul}=2\epsilon_u$, which is adequate
only near the atomic resonance, and we must adopt instead for 
$\epsilon_u$ the more general expression
\begin{equation} \label{eq:gen_width}
\epsilon_u\equiv\epsilon_u(\omega_{k'})=\frac{1}{2}\,A_{ul}\,
	\frac{\omega_{k'}^3}{\omega_{ul}^3}\;.
\end{equation}
This result will be used in Section~\ref{sec:balance}. It is worth
noting that \cite{He54} proposed that one should rather use
\begin{displaymath}
\epsilon_u\equiv\epsilon_u(\omega_{k'})=\frac{1}{2}\,A_{ul}\,
	\frac{\omega_{k'}}{\omega_{ul}}\;.
\end{displaymath}
However, \cite{Po64} pointed out that such alternative expression of the
line width is a consequence of adopting the $\bm{p\cdot A}$ form for 
the atom-photon interaction Hamiltonian (\emph{velocity gauge}), rather 
than the dipolar form $\bm{d\cdot E}$ (\emph{length gauge}; see footnote 
at p.~116 of \citealt{Po64}).
Since in this paper we consistently use the electric-dipole interaction 
Hamiltonian in the length gauge, Equation~(\ref{eq:gen_width}) represents 
the appropriate form of the line width that must be adopted in our 
treatment of radiation scattering.

Using Equations~(21)--(25), Equation~(\ref{eq:2emiss.1}) becomes
\begin{eqnarray} \label{eq:RT.JF}
\varepsilon\apx{2}_i(\omega_{k'},\bm{\hat{k}'})
&=& \frac{3}{16\pi^3}\,
	{\cal N}\hbar\,\frac{\omega_{k'}^4}{\omega_{ul}^3}\,
	\Pi_{L_u L_l}^2 A_{ul} B_{lu} \\
&&\kern -1.5cm\times
	\sum_{J_u J_u' J_u'' J_u'''}
	\sum_{J_l J_l' J_l'' J_l'''}
	\sum_{F_u F_u' F_u'' F_u'''}
	\sum_{F_l F_l' F_l'' F_l'''}
	(-1)^{J_u+J_u'+J_u''+J_u'''}
	(-1)^{J_l+J_l'+J_l''+J_l'''}
	\nonumber \\
&&\mathop{\times}
	\Pi_{J_u J_u' J_u'' J_u'''}\,
	\Pi_{J_l J_l' J_l'' J_l'''}\,
	\sixj{J_u}{J_l}{1}{L_l}{L_u}{S}
	\sixj{J_u'}{J_l'}{1}{L_l}{L_u}{S}
	\sixj{J_u''}{J_l''}{1}{L_l}{L_u}{S}
	\sixj{J_u'''}{J_l'''}{1}{L_l}{L_u}{S} \nonumber \\
&&\mathop{\times}
	\Pi_{F_u F_u' F_u'' F_u'''}\,
	\Pi_{F_l F_l' F_l'' F_l'''}\,
	\sixj{F_u}{F_l}{1}{J_l}{J_u'}{I}
	\sixj{F_u'}{F_l'}{1}{J_l'}{J_u}{I}
	\sixj{F_u''}{F_l''}{1}{J_l''}{J_u''}{I}
	\sixj{F_u'''}{F_l'''}{1}{J_l'''}{J_u'''}{I} \nonumber \\
&&\kern -1.5cm \times
	\sum_{\mu_u M_u}\sum_{\mu_u' M_u'}\sum_{\mu_l'' M_l''}
	C^{J_u F_u}_{\mu_u}(M_u)\,
	C^{J_u'' F_u''}_{\mu_u}(M_u)\,
	C^{J_u' F_u'}_{\mu_u'}(M_u')\,
	C^{J_u''' F_u'''}_{\mu_u'}(M_u')\,
	C^{J_l'' F_l''}_{\mu_l''}(M_l'')\,
	C^{J_l''' F_l'''}_{\mu_l''}(M_l'') \nonumber \\
&&\kern -1.5cm \times 
	\sum_{\bar J_l \bar J_l'}
	\sum_{\bar F_l \bar F_l'}
	\sum_{\mu_l M_l}\sum_{\mu_l' M_l'}
	C^{J_l F_l}_{\mu_l}(M_l)\,
	C^{\bar J_l \bar F_l}_{\mu_l}(M_l)\,
	C^{J_l' F_l'}_{\mu_l'}(M_l')\,
	C^{\bar J_l' \bar F_l'}_{\mu_l'}(M_l') \nonumber \\
&&\kern -1.5cm \times
	\sum_{KQ}\sum_{K'Q'}\sum_{K_l Q_l}
	\sum_{qq'} \sum_{pp'}
	(-1)^{\bar F_l-M_l+q'+p'}
	\thrj{1}{1}{K}{-q}{q'}{-Q}
	\thrj{1}{1}{K'}{-p}{p'}{-Q'}
	\thrj{\bar F_l}{\bar F_l'}{K_l}{M_l}{-M_l'}{-Q_l}
	\nonumber \\
&&\mathop{\times}
	\thrj{F_u}{F_l}{1}{-M_u}{M_l}{q}
	\thrj{F_u'}{F_l'}{1}{-M_u'}{M_l'}{q'}
	\thrj{F_u'''}{F_l'''}{1}{-M_u'}{M_l''}{p}
	\thrj{F_u''}{F_l''}{1}{-M_u}{M_l''}{p'}
	\nonumber \\
%\noalign{\eject}
&&\mathop{\times}
	\Pi_{KK'K_l}\,
	T^{K'}_{Q'}(i,\bm{\hat{k}'})\,
	\rho^{K_l}_{Q_l}(\bar J_l \bar F_l,\bar J_l' \bar F_l') 
	\nonumber \\
&&\kern -1.5cm\times 
      \sum_{j=0}^3 \oint \frac{d\bm{\hat{k}}}{4\pi}\,
	T^K_Q(j,\bm{\hat{k}})
	\int_0^\infty d\omega_k
	\left(
	\Psi_{u'l',l''ul}^{-k,+k'-k} + \bar\Psi_{ul,l''u'l'}^{-k,+k'-k}
	\right)
%	{\cal R}(\Omega_u,\Omega_{u'};
%	\Omega_l,\Omega_{l'},\Omega_{l''};
%	\omega_k,\omega_{k'})\,
	S_j(\omega_k,\bm{\hat{k}})\;,\qquad (i=0,1,2,3)
	\nonumber
\end{eqnarray}
where for simplicity of notation we removed the information of the 
term from the density matrix element and from the argument of the 
projection coefficients $C^{JF}_\mu$ of the atomic eigenstates, as 
well as defining
\begin{equation}
\Pi_{ab\ldots}\equiv\sqrt{(2a+1)(2b+1)\cdots}\;.
\end{equation}

The quantities of the first order appearing in Equation~(\ref{eq:RT.1}), 
which represent single-photon absorption and spontaneous emission, for 
the case of the multi-term atom with hyperfine structure, have been given 
by \citeauthor{CM05} 
(\citeyear{CM05}; cf.~their Equations~(32a,b) and (33a,b)). For
convenience, we reproduce those expressions using the same notation adopted 
in this paper:
\begin{eqnarray} \label{eq:emiss.JF}
\varepsilon\apx{1}_i(\omega_{k'},\bm{\hat{k}'})
&=&\frac{\sqrt3}{8\pi^2}\,
	{\cal N}\hbar\,\frac{\omega_{k'}^4}{\omega_{ul}^3}\,
	\Pi_{L_u}^2 A_{ul}
	\sum_{J_u J_u'}
	\sum_{J_l J_l'}
	\sum_{F_u F_u'}
	\sum_{F_l F_l'}
	(-1)^{J_u'-J_u+J_l'-J_l} \\
&&\mathop{\times}
	\Pi_{J_u J_u'} \Pi_{J_l J_l'}
	\Pi_{F_u F_u'} \Pi_{F_l F_l'}\,
	\sixj{J_u}{J_l}{1}{L_l}{L_u}{S}
	\sixj{J_u'}{J_l'}{1}{L_l}{L_u}{S}
	\sixj{F_u}{F_l}{1}{J_l}{J_u}{I}
	\sixj{F_u'}{F_l'}{1}{J_l'}{J_u'}{I}
	\nonumber \\
&&\kern -1.5cm\times
	\sum_{\bar J_u \bar J_u'}
	\sum_{\bar F_u \bar F_u'}
	\sum_{\mu_u M_u}\sum_{\mu_u' M_u'}\sum_{\mu_l M_l}
	C^{J_u F_u}_{\mu_u}(M_u)\,
	C^{\bar J_u \bar F_u}_{\mu_u}(M_u)\,
	C^{J_u' F_u'}_{\mu_u'}(M_u')\,
	C^{\bar J_u' \bar F_u'}_{\mu_u'}(M_u')\,
	C^{J_l F_l}_{\mu_l}(M_l)\,
	C^{J_l' F_l'}_{\mu_l}(M_l) \nonumber \\
\noalign{\eject}
&&\kern -1.5cm\times
	\sum_{KQ}\sum_{K_u Q_u}
	\sum_{qq'}
	(-1)^{\bar F_u-M_u'+q'+1}
	\thrj{1}{1}{K}{-q}{q'}{-Q}\,
	\thrj{\bar F_u}{\bar F_u'}{K_u}{M_u}{-M_u'}{-Q_u}
	\thrj{F_u'}{F_l'}{1}{-M_u'}{M_l}{q}
	\thrj{F_u}{F_l}{1}{-M_u}{M_l}{q'}
	\nonumber \\
&&\mathop{\times}
	\Pi_{K K_u}\,
	T^K_Q(i,\bm{\hat{k}'})\,
	\rho^{K_u}_{Q_u}(\bar J_u \bar F_u,\bar J_u' \bar F_u') 
	\left(
	\Phi_{lu'}^{+k'} + \bar\Phi_{lu}^{+k'}
	\right)\;,\qquad (i=0,1,2,3) \nonumber
\end{eqnarray}
\begin{eqnarray} \label{eq:absorb.JF}
\eta_i(\omega_{k'},\bm{\hat{k}'})
&=&\frac{\sqrt3}{8\pi^2}\,
	{\cal N}\hbar\omega_{k'}\,
	\Pi_{L_l}^2 B_{lu}
	\sum_{J_u J_u'}
	\sum_{J_l J_l'}
	\sum_{F_u F_u'}
	\sum_{F_l F_l'}
	(-1)^{J_u'-J_u+J_l'-J_l} \\
&&\mathop{\times}
	\Pi_{J_u J_u'} \Pi_{J_l J_l'}
	\Pi_{F_u F_u'} \Pi_{F_l F_l'}\,
	\sixj{J_u}{J_l}{1}{L_l}{L_u}{S}
	\sixj{J_u'}{J_l'}{1}{L_l}{L_u}{S}
	\sixj{F_u}{F_l}{1}{J_l}{J_u}{I}
	\sixj{F_u'}{F_l'}{1}{J_l'}{J_u'}{I}
	\nonumber \\
&&\kern -1.5cm\times
	\sum_{\bar J_l \bar J_l'}
	\sum_{\bar F_l \bar F_l'}
	\sum_{\mu_l M_l}\sum_{\mu_l' M_l'}\sum_{\mu_u M_u}
	C^{J_l F_l}_{\mu_l}(M_l)\,
	C^{\bar J_l \bar F_l}_{\mu_l}(M_l)\,
	C^{J_l' F_l'}_{\mu_l'}(M_l')\,
	C^{\bar J_l' \bar F_l'}_{\mu_l'}(M_l')\,
	C^{J_u F_u}_{\mu_u}(M_u)\,
	C^{J_u' F_u'}_{\mu_u}(M_u) \nonumber \\
&&\kern -1.5cm\times
	\sum_{KQ}\sum_{K_l Q_l}
	\sum_{qq'}
	(-1)^{\bar F_l-M_l+q'+1}
	\thrj{1}{1}{K}{-q}{q'}{-Q}\,
	\thrj{\bar F_l}{\bar F_l'}{K_l}{M_l}{-M_l'}{-Q_l}
	\thrj{F_u}{F_l}{1}{-M_u}{M_l}{q}
	\thrj{F_u'}{F_l'}{1}{-M_u}{M_l'}{q'}
	\nonumber \\
&&\mathop{\times}
	\Pi_{K K_l}\,
	T^K_Q(i,\bm{\hat{k}'})\,
	\rho^{K_l}_{Q_l}(\bar J_l \bar F_l,\bar J_l' \bar F_l') 
	\left(
	\Phi_{ul'}^{-k'} + \bar\Phi_{ul}^{-k'}
	\right)\;,\qquad (i=0,1,2,3) \nonumber
\end{eqnarray}
having also defined
\begin{equation} \label{eq:Phi}
\Phi_{ab}^{\pm k} 
\equiv \frac{\rm i}{\omega_{ba}\mp\omega_k+{\rm i}\epsilon_{ba}}\;,
\end{equation}
where $\ket{a}$ and $\ket{b}$ are in general eigenstates of 
the atomic Hamiltonian including the contributions of the external 
fields. Again, we used the notation $\bar{\Phi}$ to indicate the
complex conjugate of the profile $\Phi$.

The magneto-optical coefficients $\rho_i(\omega_{k'},\bm{\hat{k}'})$, 
for $i=1,2,3$ \cite[see Equation~(32a) of][]{CM05}, are obtained 
from the expression of $\eta_i(\omega_{k'},\bm{\hat{k}'})$,
Equation~(\ref{eq:absorb.JF}), through the following substitution,
\begin{displaymath}
	\left(
	\Phi_{ul'}^{-k'} + \bar\Phi_{ul}^{-k'}
	\right) \longrightarrow
	{\rm i} \left(
	\Phi_{ul'}^{-k'} - \bar\Phi_{ul}^{-k'}
	\right)\;.
\end{displaymath}

The elements of the absorption matrix 
$\kappa_{ij}(\omega_{k'},\bm{\hat{k}'})$, which appear in
Equation~(\ref{eq:RT.1}), are related to the coefficients 
$\eta_i(\omega_{k'},\bm{\hat{k}'})$ and 
$\rho_i(\omega_{k'},\bm{\hat{k}'})$ via the following relations
\begin{gather}
\label{eq:dichroic}
\kappa_{ii}=\eta_0\;,\qquad
\kappa_{0i}=\kappa_{i0}=\eta_i\;,\qquad(i=0,1,2,3) \\
\label{eq:magnetooptical}
\kappa_{ij}=\epsilon_{ijk}\,\rho_k\;.\qquad(i,j,k=1,2,3)
\end{gather}
where $\epsilon_{ijk}$ is the fully antisymmetric Levi-Civita tensor.

As a special case, we consider the limit of zero magnetic field for
Equations~(\ref{eq:RT.JF}), (\ref{eq:emiss.JF}), and (\ref{eq:absorb.JF}). 
In that case, there is no dependence of the line profiles 
%$\left(
%\Psi_{ul',l''u'l}^{-k,+k'-k} + \bar\Psi_{u'l,l''ul'}^{-k,+k'-k}
%\right)$
%the argument of 
%${\cal R}(\Omega_u,\Omega_{u'}; \Omega_l,\Omega_{l'},\Omega_{l''};
%	\omega_k,\omega_{k'})$ 
on the $\mu$-indices, and so
we can use the orthogonality properties (\ref{eq:diag.prop}) to
perform the trivial summations over those indices. In addition, it is 
possible to sum over all magnetic quantum numbers, as well as over the 
indices $(q,q')$ and $(p,p')$. After some tedious Racah-algebra
manipulations \cite[e.g.,][]{Va88}, we find
\begin{eqnarray}	\label{eq:RT.JF.noB}
\varepsilon\apx{2}_i(\omega_{k'},\bm{\hat{k}'})\biggr|_{B=0}
&=& \frac{3}{16\pi^3}\,
	{\cal N}\hbar\,\frac{\omega_{k'}^4}{\omega_{ul}^3}\,
	\Pi_{L_u L_l}^2 A_{ul} B_{lu}
	\sum_{J_l J_l'J_l''}
	\sum_{J_u J_u'}
	\sum_{F_l F_l'F_l''}
	\sum_{F_u F_u'}
	(-1)^{F_u'+F_l''+1}\,
	\\
%\noalign{\eject}
&&\mathop{\times}
	\Pi_{J_l J_l'}
	\Pi_{J_l''J_u J_u'}^2
	\sixj{J_u}{J_l}{1}{L_l}{L_u}{S}
	\sixj{J_u'}{J_l'}{1}{L_l}{L_u}{S}
	\sixj{J_u}{J_l''}{1}{L_l}{L_u}{S}
	\sixj{J_u'}{J_l''}{1}{L_l}{L_u}{S}
	\nonumber \\
&&\mathop{\times}
	\Pi_{F_l F_l'}
	\Pi_{F_l''F_u F_u'}^2 
	\sixj{F_u}{F_l}{1}{J_l}{J_u}{I}
	\sixj{F_u'}{F_l'}{1}{J_l'}{J_u'}{I}
	\sixj{F_u}{F_l''}{1}{J_l''}{J_u}{I}
	\sixj{F_u'}{F_l''}{1}{J_l''}{J_u'}{I}
	\nonumber \\
&&\kern -2.5cm\times
	\sum_{KQ}\sum_{K'Q'}\sum_{K_l Q_l} 
	(-1)^{K_l-Q_l}\,\Pi_{KK'K_l}
	\thrj{K}{K'}{K_l}{Q}{Q'}{-Q_l}
	\sixj{K'}{F_u}{F_u'}{F_l''}{1}{1}
	\ninj{K}{K'}{K_l}{1}{F_u'}{F_l'}{1}{F_u}{F_l}
	\nonumber \\
\noalign{\vspace{-3pt}}
&&\mathop{\times}
%	\sum_x \Pi_x^2\,
%	\sixj{K}{F_u}{x}{F_l'}{1}{1}
%	\sixj{F_l'}{1}{x}{F_u'}{K_l}{F_l}
%	\sixj{F_u'}{K_l}{x}{K}{F_u}{K'}
	T^{K'}_{Q'}(i,\bm{\hat{k}'})\,  
	\rho^{K_l}_{Q_l}(J_l F_l,J_l'F_l')
	\nonumber \\
&&\kern -2.5cm\times 
      \sum_{j=0}^3 \oint \frac{d\bm{\hat{k}}}{4\pi}\,
	T^K_Q(j,\bm{\hat{k}})
	\int_0^\infty d\omega_k
	\left(
	\Psi_{u'l',l''ul}^{-k,+k'-k} + \bar\Psi_{ul,l''u'l'}^{-k,+k'-k}
	\right)
%	{\cal R}(\Omega_u,\Omega_{u'};
%	\Omega_l,\Omega_{l'},\Omega_{l''};
%	\omega_k,\omega_{k'})\,
	S_j(\omega_k,\bm{\hat{k}})\;,\qquad (i=0,1,2,3)
	\nonumber
\end{eqnarray}
whereas the contributions of the first-order terms for this 
special case of vanishing magnetic fields are given by\footnote{%
We take here the opportunity to correct a statement given by \cite{CM05}. 
Equations~(34) in
that paper are only valid when both the fine and hyperfine structures 
of the levels -- and not just the Zeeman splitting, as originally 
claimed in the paper -- can also be neglected with respect to the width 
of the line profile. In particular, those equations hold for the 
frequency-integrated polarization. The case of zero magnetic field is 
instead properly represented by Equations~(\ref{eq:emiss1}) and 
(\ref{eq:absorb1}) given here.}
\begin{eqnarray} \label{eq:emiss1}
\varepsilon\apx{1}_i(\omega_{k'},\bm{\hat{k}'})\biggr|_{B=0}
&=&\frac{\sqrt3}{8\pi^2}\,
	{\cal N}\hbar\,\frac{\omega_{k'}^4}{\omega_{ul}^3}\,
	\Pi_{L_u}^2 A_{ul}
	\sum_{J_u J_u'}
	\sum_{F_u F_u'}
	\sum_{J_l F_l}
	(-1)^{J_u-J_u'+F_u+F_l+1}\,
	\Pi_{J_u J_u'F_u F_u'}\Pi_{J_l F_l}^2 \\
&&\mathop{\times}
	\sixj{J_u}{J_l}{1}{L_l}{L_u}{S}
	\sixj{J_u'}{J_l}{1}{L_l}{L_u}{S}
	\sixj{F_u}{F_l}{1}{J_l}{J_u}{I}
	\sixj{F_u'}{F_l}{1}{J_l}{J_u'}{I}
	\nonumber \\
&&\kern -1cm\times
	\sum_{KQ}\,
	\sixj{K}{F_u}{F_u'}{F_l}{1}{1}\;
	T^K_Q(i,\bm{\hat{k}'})\,
	\rho^K_Q(J_u F_u,J_u'F_u')
	\left(\Phi^{+k'}_{lu'}+\bar\Phi^{+k'}_{lu}\right)\;,
	\qquad (i=0,1,2,3) \nonumber
\end{eqnarray}
\begin{eqnarray} \label{eq:absorb1}
\eta_i(\omega_{k'},\bm{\hat{k}'})\biggr|_{B=0}
&=&\frac{\sqrt3}{8\pi^2}\,
	{\cal N}\hbar\omega_{k'}\,
	\Pi_{L_l}^2 B_{lu}
	\sum_{J_l J_l'}
	\sum_{F_l F_l'}
	\sum_{J_u F_u}
	(-1)^{J_l'-J_l+F_u+F_l+1}\,
	\Pi_{J_l J_l'F_l F_l'}\Pi_{J_u F_u}^2 \\
&&\mathop{\times}
	\sixj{J_u}{J_l}{1}{L_l}{L_u}{S}
	\sixj{J_u}{J_l'}{1}{L_l}{L_u}{S}
	\sixj{F_u}{F_l}{1}{J_l}{J_u}{I}
	\sixj{F_u}{F_l'}{1}{J_l'}{J_u}{I}
	\nonumber \\
&&\kern -1cm\times
	\sum_{KQ}(-1)^K\,
	\sixj{K}{F_l}{F_l'}{F_u}{1}{1}\;
	T^K_Q(i,\bm{\hat{k}'})\,
	\rho^K_Q(J_l F_l,J_l'F_l')
	\left(\Phi^{-k'}_{ul'}+\bar\Phi^{-k'}_{ul}\right)\;.
	\qquad (i=0,1,2,3) \nonumber
\end{eqnarray}

In the absence of hyperfine structure ($I=0$), Equation~(\ref{eq:RT.JF.noB})
reduces to
\begin{eqnarray}
\varepsilon\apx{2}_i(\omega_{k'},\bm{\hat{k}'})\biggr|_{B=0}
&=&\frac{3}{16\pi^3}\,
	{\cal N}\hbar\,\frac{\omega_{k'}^4}{\omega_{ul}^3}\,
	\Pi_{L_u L_l}^2 A_{ul} B_{lu} \\
&&\kern -1.8cm\times
	\sum_{J_l J_l' J_l''}
	\sum_{J_u J_u'}
	(-1)^{J_u'+J_l''+1}\,
	\Pi_{J_l J_l'}
	\Pi_{J_l''J_u J_u'}^2
	\sixj{J_u}{J_l}{1}{L_l}{L_u}{S}
	\sixj{J_u'}{J_l'}{1}{L_l}{L_u}{S}
	\sixj{J_u}{J_l''}{1}{L_l}{L_u}{S}
	\sixj{J_u'}{J_l''}{1}{L_l}{L_u}{S}
	\nonumber \\
&&\kern -1.8cm\times
	\sum_{KQ}\sum_{K'Q'}\sum_{K_l Q_l} 
	(-1)^{K_l-Q_l}\,\Pi_{KK'K_l}
	\thrj{K}{K'}{K_l}{Q}{Q'}{-Q_l}
	\sixj{K'}{J_u}{J_u'}{J_l''}{1}{1}
	\ninj{K}{K'}{K_l}{1}{J_u'}{J_l'}{1}{J_u}{J_l}
	\nonumber \\
&&\mathop{\times}
%	\sum_x \Pi_x^2\,
%	\sixj{K}{J_u}{x}{J_l'}{1}{1}
%	\sixj{J_l'}{1}{x}{J_u'}{K_l}{J_l}
%	\sixj{J_u'}{K_l}{x}{K}{J_u}{K'}
	T^{K'}_{Q'}(i,\bm{\hat{k}'})\,
	\rho^{K_l}_{Q_l}(J_l,J_l') \nonumber \\
&&\kern -1.8cm\times 
      \sum_{j=0}^3 \oint \frac{d\bm{\hat{k}}}{4\pi}\,
	T^K_Q(j,\bm{\hat{k}})
	\int_0^\infty d\omega_k
	\left(
	\Psi_{u'l',l''ul}^{-k,+k'-k} + \bar\Psi_{ul,l''u'l'}^{-k,+k'-k}
	\right)
%	{\cal R}(\Omega_u,\Omega_{u'};
%	\Omega_l,\Omega_{l'},\Omega_{l''};
%	\omega_k,\omega_{k'})\,
	S_j(\omega_k,\bm{\hat{k}})\;,\qquad (i=0,1,2,3)
	\nonumber
\end{eqnarray}
and finally, in the additional case of unpolarized lower term
(u.l.s.: $K_l=0$, $J_l'=J_l$), 
\begin{eqnarray} \label{eq:RT.J.metalevel}
\varepsilon\apx{2}_i(\omega_{k'},\bm{\hat{k}'})\biggr|_{B=0}^\textrm{u.l.s.}
&=&\frac{3}{16\pi^3}\,
	{\cal N}\hbar\,\frac{\omega_{k'}^4}{\omega_{ul}^3}\,
	\Pi_{L_u L_l}^2 A_{ul} B_{lu} \\
&&\kern -2.1cm\times 
	\sum_{J_l J_l'}
	\sum_{J_u J_u'}
	(-1)^{J_l-J_l'}\,
	\frac{%
	\Pi_{J_l J_l'J_u J_u'}^2
	}{%
	\epsilon_{J_u J_u'}+{\rm i}\omega_{J_u J_u'}
	}\,
	\sixj{J_u}{J_l}{1}{L_l}{L_u}{S}
	\sixj{J_u'}{J_l'}{1}{L_l}{L_u}{S}
	\sixj{J_u'}{J_l}{1}{L_l}{L_u}{S}
	\sixj{J_u}{J_l'}{1}{L_l}{L_u}{S}
	\nonumber \\
&&\kern -2.1cm\times 
	\sum_{KQ} (-1)^Q\,
	\sixj{K}{J_u}{J_u'}{J_l}{1}{1}
	\sixj{K}{J_u}{J_u'}{J_l'}{1}{1}\,
	\frac{1}{\sqrt{2J_l+1}}\,
	T^K_{-Q}(i,\bm{\hat{k}'})\,
	\rho^{0}_{0}(J_l,J_l) \nonumber \\
&&\kern -2.1cm\times 
      \sum_{j=0}^3 \oint \frac{d\bm{\hat{k}}}{4\pi}\,
	T^K_Q(j,\bm{\hat{k}})
	\int_0^\infty d\omega_k\,
	{\cal R}(\Omega_u,\Omega_{u'};
	\Omega_l,\Omega_{l'};
	\omega_k,\omega_{k'})_\textrm{n.c.l.t.}\,
	S_j(\omega_k,\bm{\hat{k}})\;,\qquad (i=0,1,2,3)
	\nonumber
\end{eqnarray}
where we recalled Equation~(\ref{eq:red.nclt}), and renamed $l''$ into $l'$.

In the limit of sharp lower levels, this last expression can be compared 
with the one for the redistribution matrix derived for the same model 
atom by \cite{La97}, using the metalevel approach to the description
of the atomic density matrix. We note that the redistribution function 
${\cal R}(\Omega_u,\Omega_{u'};\Omega_l,\Omega_{l'};
	\omega_k,\omega_{k'})_\textrm{n.c.l.t.}$ in the integral
must be replaced by the function ${\cal R}_2$ of Equation~(\ref{eq:red.sll}) 
for this comparison. Once this substitution is made, and taking into 
account the vanishing of the 1st-order emissivity in the limit of sharp 
lower levels, 
we can demonstrate the formal identity of Equation~(\ref{eq:RT.J.metalevel}) 
with the result of that former derivation.

Similarly, imposing $S=I=0$ in Equation~(\ref{eq:RT.JF}), we can derive 
the coherent scattering emissivity for the two-level atom in the
presence of a magnetic field. This also turns out to be identical with the 
analogous result derived by \cite{La97}, in the limit of sharp and 
unpolarized lower level considered by those authors. 
%
%After some straightforward Racah algebra, we find
%%
%\begin{eqnarray} \label{eq:RT.J.B.metalevel}
%\varepsilon\apx{2}_i(\omega_{k'},\bm{\hat{k}'})
%&=& \frac{3}{16\pi^3}\,
%	{\cal N}\hbar\,\frac{\omega_{k'}^4}{\omega_{ul}^3}\,
%	\Pi_{J_u}^2 A_{ul} B_{lu}\,
%	\sqrt{2J_l+1}\,
%	\rho^{0}_{0}(J_l,J_l) \\
%&&\kern -1.5cm \times
%	\sum_{M_u M_u'}
%	\sum_{KQ}\sum_{K'Q'}
%	\sum_{qq'} \sum_{pp'}
%	(-1)^{q'+p'}
%	\frac{\Pi_{KK'}}{2\epsilon_u+{\rm i}\omega_{uu'}}\,
%	\thrj{1}{1}{K}{-q}{q'}{-Q}
%	\thrj{1}{1}{K'}{-p}{p'}{-Q'}\,
%	T^{K'}_{Q'}(i,\bm{\hat{k}'})
%	\nonumber \\
%&&\mathop{\times}
%	\sum_{M_l M_l'}
%	\thrj{J_u}{J_l}{1}{-M_u}{M_l}{q}
%	\thrj{J_u}{J_l}{1}{-M_u'}{M_l}{q'}
%	\thrj{J_u}{J_l}{1}{-M_u'}{M_l'}{p}
%	\thrj{J_u}{J_l}{1}{-M_u}{M_l'}{p'}
%	\nonumber \\
%&&\kern -1.5cm\times 
%      \sum_{j=0}^3 \oint \frac{d\bm{\hat{k}}}{4\pi}\,
%	T^K_Q(j,\bm{\hat{k}})
%	\int_0^\infty d\omega_k\,
%	{\cal R}_2(\Omega_u,\Omega_{u'};
%	\Omega_l,\Omega_{l'};
%	\omega_k,\omega_{k'})\,
%	S_j(\omega_k,\bm{\hat{k}})\;,\qquad (i=0,1,2,3)
%	\nonumber
%\end{eqnarray}
%
%which also is found to agree with the metalevel result.

\section{Verification of the energy-balance condition for coherent scattering}
\label{sec:balance}

A fundamental property of the transport of radiation in a 
collisionless gas of atoms at the statistical equilibrium (i.e., 
under stationary conditions, 
such that the internal energy of the atomic ensemble does not change 
with time) is that the radiation's energy flux through a closed surface 
containing the gas must be zero. It is thus important to 
verify that 
Equation~(\ref{eq:RT.1}) --  which generalizes the usual radiative transfer
equation for polarized radiation \cite[e.g.,][]{LL04} via the inclusion 
of coherent scattering -- satisfies such fundamental energy-balance 
condition.

%For simplicity, we consider the case of an atom with two levels,
%respectively with angular momentum $J_l$ and $J_u$. We can then use
%the equations of the previous section assuming $S=I=0$. 
%
%We also assume that the 
%incident radiation is described by a collimated beam 
%of unpolarized and monochromatic light propagating along the $z$-axis,
% and of frequency $\omega_0$,
%%
%\begin{displaymath}
%S_i(\omega_k,\bm{\hat{k}})
%	=\delta_{i0}\,\delta(\bm{\hat{k}}-\bm{\hat{z}})\,
%%	 \delta(\omega_k-\omega_0)\,{\cal S}_0(\omega_0)\;.
%S_0(\omega_k,\bm{\hat{z}})\;.
%\end{displaymath}
%%
Because our present theory cannot account for the broadening of the 
lower levels due to photon absorption
(see discussion at the end of Section~\ref{sec:lifetimes}), 
%these transitions can only be virtual, in the absence of collisions. 
%Thus, in order to guarantee the self-consistency of this verification, 
in order to guarantee the self-consistency of this verification, we 
only consider the case of infinitely sharp lower levels. 
It must be noted that this choice is consistent with the approximation 
of neglecting stimulated emission. In fact, if the incident radiation 
field is strong enough to induce any appreciable broadening of the lower 
levels, then one should also expect that stimulation effects will be 
comparatively important, and therefore should be included in the picture. 
This is certainly the case when the objective is to verify the energy balance 
among the various radiation processes.

As a consequence of the assumption of infinite lifetime of the lower
level, we then must drop the contribution of the 1st-order 
emissivity to the radiative energy balance. Nonetheless, we will retain a 
finite width of the lower levels in the expressions of the line profiles, 
for the sake of this demonstration.
This is because, in real cases, these levels may still be broadened by 
collisional processes, or by the possibility of spontaneous de-excitation 
towards other lower terms, in the case of subordinate lines in a 
multi-term atom.

The verification of the radiative balance condition in the collisionless 
regime can then be set up in general form starting from 
Equation~(\ref{eq:RT.1}). For simplicity, we assume that the incident 
radiation is unpolarized,
\begin{displaymath}
S_i(\omega_k,\bm{\hat{k}})
	=\delta_{i0}\,S_0(\omega_k,\bm{\hat{k}})\;.
\end{displaymath}
We then can ignore the contribution from the absorption terms associated 
with the magneto-optical effects (see
Equation~(\ref{eq:magnetooptical})), and only need to consider
the expression of the absorption coefficient adopting the same 
notations used in writing Equation~(\ref{eq:2emiss.1}),
\begin{eqnarray} \label{eq:absorb.gen}
\eta_i(\omega_{k'},\bm{\hat{k}'})
&=& \frac{2\pi}{\sqrt3}\,\frac{e_0^2}{\hbar c}\,
	{\cal N}\omega_{k'}
	\sum_{ll'}\rho_{ll'}\sum_u\sum_{qq'}(-1)^{q'+1}\,
	(r_q)_{ul}(r_{q'})_{ul'}^* \\
&&{}\times
	\sum_{KQ}\sqrt{2K+1}\,\thrj{1}{1}{K}{-q}{q'}{-Q}\,
	T^K_Q(i,\bm{\hat{k}'})
	\left(
	\Phi_{ul'}^{-k'} + \bar\Phi_{ul}^{-k'}
	\right)\;.\qquad (i=0,1,2,3) \nonumber
\end{eqnarray}

Using the tabulated expressions for
$T^K_Q(i,\bm{\hat{k}})$ \citep{Bo97b,LL04}, we can show that
%
%\begin{eqnarray}
\begin{equation}
\label{eq:avg1}
\oint\frac{d\bm{\hat{k}}}{4\pi}\,
	T^K_Q(i,\bm{\hat{k}})
%&=&\delta_{Q0} \left(\delta_{i0}\,\delta_{K0}
=\delta_{Q0} \left(\delta_{i0}\,\delta_{K0}
	-\frac{1}{\sqrt{2}}\,
	\delta_{i1}\,\delta_{K2}\right)\;.% \\
%\label{eq:avg2}
%\oint\frac{d\bm{\hat{k}}}{4\pi}\,
%	T^K_Q(i,\bm{\hat{k}})\,S_j(\omega_k,\bm{\hat{k}})
%&=&\delta_{j0}\,\delta_{Q0} \left[
%	\delta_{i0}\left(\delta_{K0}+\frac{1}{\sqrt{2}}\,\delta_{K2}\right)
%	+\delta_{i3}\,\delta_{K1}\sqrt\frac{3}{2} \right]
%S_0(\omega_k,\bm{\hat{z}})\;.
%%	\delta(\omega_k-\omega_0)\,{\cal S}_0(\omega_0)\;.\kern .5cm
%\end{eqnarray}
\end{equation}
With the above assumptions, and for $i=0$, from
Equations~(\ref{eq:RT.1}), (\ref{eq:2emiss.1}), (\ref{eq:dichroic}),
and (\ref{eq:absorb.gen}), after some straightforward algebra, we find
\begin{eqnarray} \label{eq:balance}
&&\oint\frac{d\bm{\hat{k}}}{4\pi}
\int_0^\infty d\omega_{k}\,
\frac{1}{c}\,\frac{d}{dt}\,S_0(\omega_{k},\bm{\hat{k}}) \\
&=&{}-\frac{2\pi}{\sqrt3}\,\frac{e_0^2}{\hbar c}\,
	{\cal N}
	\sum_{ll'}\rho_{ll'}
	\sum_{KQ}\sum_{qq'}(-1)^{q'+1}\,
	\Pi_K\,\thrj{1}{1}{K}{-q}{q'}{-Q}
	\oint\frac{d\bm{\hat{k}}}{4\pi}\,T^K_Q(0,\bm{\hat{k}})
	\int_0^\infty d\omega_{k}\,S_0(\omega_k,\bm{\hat{k}}) \nonumber \\
&&{}\times\Biggl\{
	\sum_u (r_q)_{ul}(r_{q'})_{ul'}^*\,
	\omega_k \left(
	\Phi_{ul'}^{-k} + \bar\Phi_{ul}^{-k}
	\right) \nonumber \\
&&\hphantom{{}\times\Biggl\{} -
	\frac{2}{3}\frac{e_0^2}{\hbar c^3}
	\sum_{uu'} (r_q)_{ul}(r_{q'})_{u'l'}^*
	\sum_{pl''} (r_p)_{u'l''}(r_p)_{ul''}^*\,
	\frac{1}{\pi}
	\int_0^\infty d\omega_{k'}\,\omega_{k'}^4
	\left(
	\Psi_{u'l',l''ul}^{-k,+k'-k} + \bar\Psi_{ul,l''u'l'}^{-k,+k'-k}
	\right) \Biggr\} \nonumber \\
%\noalign{\eject}
&\equiv&{}-\frac{2\pi}{\sqrt3}\,\frac{e_0^2}{\hbar c}\,
	{\cal N}
	\sum_{ll'}\rho_{ll'}
	\sum_{KQ}\sum_{qq'}(-1)^{q'+1}\,
	\Pi_K\,\thrj{1}{1}{K}{-q}{q'}{-Q}
	\oint\frac{d\bm{\hat{k}}}{4\pi}\,T^K_Q(0,\bm{\hat{k}})
	\int_0^\infty d\omega_{k}\,S_0(\omega_k,\bm{\hat{k}}) \nonumber \\
&&{}\times\Biggl\{
	\sum_u (r_q)_{ul}(r_{q'})_{ul'}^*\,
	\omega_k \left(
	\Phi_{ul'}^{-k} + \bar\Phi_{ul}^{-k}
	\right) \nonumber \\
&&\hphantom{{}\times\Biggl\{} -
%	\frac{2}{3}\frac{e_0^2}{\hbar c^3}
	\sum_{uu'} (r_q)_{ul}(r_{q'})_{u'l'}^*
	\frac{ \sum_{pl''} (r_p)_{u'l''}(r_p)_{ul''}^* }
	     {\sum_{p\bar{l}} |(r_p)_{u\bar{l}}|^2 }\,
	\frac{1}{\pi}
	\int_0^\infty d\omega_{k'}\;\epsilon_u(\omega_{k'})\,\omega_{k'}
	\left(
	\Psi_{u'l',l''ul}^{-k,+k'-k} + \bar\Psi_{ul,l''u'l'}^{-k,+k'-k}
	\right) \Biggr\}\;, \nonumber
\end{eqnarray}
where in the last equivalence we used the definition of
$\epsilon_u(\omega_{k'})$, Equations~(\ref{eq:gen_width}) and 
(\ref{eq:Einstein A}). %(see also Equation~(??) of \citealt{La97}). 
In the following, we neglect the (weak) dependence of 
$\epsilon_u(\omega_{k'})$ on the frequency of the scattered radiation,
so that it can be taken outside the integral in the last line of
Equation~(\ref{eq:balance}).
This is justified by the fact that the redistribution function appearing
in that integral effectively limits the frequency range around the 
resonance frequency $\omega_{ul}$ that contributes to the finite value 
of the integral.

From Equation~(\ref{eq:red.gen}), and using the form of the redistribution 
function given in Equation~(\ref{eq:red.gen.alt}), we see that the 
integration over 
$\omega_{k'}$ in Equation~(\ref{eq:balance}) implies the evaluation of 
four divergent integrals. In order to accomplish this task, we apply a 
consistent cut-off procedure, which is derived from the usual Feynman's 
regularization techniques of quantum field theory \cite[e.g.,][]{BS59}. 
We thus find, once again extending the integration domain to $-\infty$
(see Note~\ref{note:domainextend})
\begin{eqnarray} \label{eq:integral}
&&\kern -1cm
\frac{1}{\pi}
\int_{-\infty}^\infty d\omega_{k'}\,\omega_{k'}
\left(
\Psi_{u'l',l''ul}^{-k,+k'-k} + \bar\Psi_{ul,l''u'l'}^{-k,+k'-k}
\right) \nonumber \\
&=& 
	\frac{\omega_k+\omega_{ll''}-{\rm i}\epsilon_{ll''}}{%
	(\omega_k-\omega_{ul}+{\rm i}\epsilon_{u}-{\rm i}\epsilon_{l})
	(\omega_k-\omega_{u'l}-{\rm i}\epsilon_{u'l}) } +
	\frac{\omega_k+\omega_{l'l''}+{\rm i}\epsilon_{l'l''}}{%
	(\omega_k-\omega_{ul'}+{\rm i}\epsilon_{ul'})
	(\omega_k-\omega_{u'l'}-{\rm i}\epsilon_{u'}+{\rm i}\epsilon_{l'}) } 
	\nonumber \\
&+&\frac{\epsilon_{ll'}+{\rm i}\,\omega_{ll'}} 
	  {\epsilon_{uu'}+{\rm i}\,\omega_{uu'}} \biggl[
	\frac{\omega_{ul''}-{\rm i}\epsilon_{ul''}}{%
	(\omega_k-\omega_{ul}+{\rm i}\epsilon_{u}-{\rm i}\epsilon_{l})
	(\omega_k-\omega_{ul'}+{\rm i}\epsilon_{ul'}) } \nonumber \\
&&\hphantom{{}+\frac{\epsilon_{ll'}+{\rm i}\,\omega_{ll'}} 
	  {\epsilon_{uu'}+{\rm i}\,\omega_{uu'}} \biggl[ }+
	\frac{\omega_{u'l''}+{\rm i}\epsilon_{u'l''}}{%
	(\omega_k-\omega_{u'l}-{\rm i}\epsilon_{u'l})
	(\omega_k-\omega_{u'l'}-{\rm i}\epsilon_{u'}+{\rm i}\epsilon_{l'}) } 
	\biggr]\;.
\end{eqnarray}
We note that the dependence of this integral on $\omega_{l''}$ is
confined to the numerator. We then introduce an approximation, 
which is to replace $\omega_{l''}$ in Equation~(\ref{eq:integral})
by some appropriate average $\bar{\omega}_l$. 
If we further assume that the conditions for the diagonality of the
damping matrix $\mathit{\Gamma}_u$ apply (see Section~\ref{sec:lifetimes}), 
we are then able to formally perform the summation over $p$ and $l''$ in
Equation~(\ref{eq:balance}),
\begin{equation} \label{eq:last.result1}
\sum_{pl''} (r_p)_{u'l''}(r_p)_{ul''}^*
	=\delta_{uu'} \sum_{pl''} |(r_p)_{ul''}|^2\;.
\end{equation}
Under these assumptions, after some more tedious algebra, we obtain
\begin{eqnarray} \label{eq:last.result2}
\frac{\epsilon_u}{\pi}
\int_{-\infty}^\infty d\omega_{k'}\,\omega_{k'}
\left(
\Psi_{ul',l''ul}^{-k,+k'-k} + \bar\Psi_{ul,l''ul'}^{-k,+k'-k}
\right)
&=& \frac{\epsilon_u+{\rm i}\,(\omega_u-\bar{\omega}_l)}{%
	\omega_k-\omega_{ul'}+{\rm i}\epsilon_{ul'}} +
  \frac{\epsilon_u-{\rm i}\,(\omega_u-\bar{\omega}_l)}{%
	\omega_k-\omega_{ul}-{\rm i}\epsilon_{ul}} \nonumber \\
&\approx& 
	\omega_k \left(
	\Phi_{ul'}^{-k} + \bar\Phi_{ul}^{-k}
	\right)\;,
\end{eqnarray}
where, for the last approximation, we took into account the hypothesis 
of sharp lower levels, and recalled the definition of the profiles 
$\Phi^{\pm k}_{ab}$, Equation~(\ref{eq:Phi}).
%For the last passage of 
%Equation~(\ref{eq:last.result2}), we noted that 
%%$\epsilon_u\ll\omega_u-\bar{\omega}_l$ in most cases, so we could drop 
%%$\epsilon_u$ from the numerators. We also considered that 
%$\omega_k\approx\omega_u-\bar{\omega}_l\pm{\rm i}\epsilon_u$, when the 
%above expression appears in an integral over $\omega_k$, like in
%Equation~(\ref{eq:balance}) (cf.~also Equation~(\ref{eq:integral})).
%
In particular, \emph{this approximation becomes exact in the absence of
lower-level polarization, or if the magnetic field is zero.}
Using then the results of Equations~(\ref{eq:last.result1}) and
(\ref{eq:last.result2}), the quantity within curly braces in 
Equation~(\ref{eq:balance}) vanishes, thus verifying the expected 
condition of radiative balance in a collisionless gas of atoms.

It is important to note that the approximations involved in this
verification are dominated by the assumption that the energy separation
among the lower levels $\{l,l',l'',\ldots\}$ is much smaller than the
frequency of the incident radiation, $\omega_k$, so that the energy values
of those levels can all be approximated by the average $\bar{\omega}_l$. 
In the presence of a magnetic field, the energy separation of the atomic 
levels is proportional to the Larmor frequency for the applied field, 
$\omega_B$. 
For typical observing conditions in the solar atmosphere, the ratio 
$\omega_B/\omega_k$ is of the order of $10^{-4}$ or less 
(e.g., for $\lambda\sim 1\,\rm\mu m$, and $B\sim 10^3\,\rm G$,
$\omega_B/\omega_k\approx 5\times 10^{-6}$), so the above approximation is
well justified.
For transitions in complex atoms, the energy span of the various 
$\omega_{l''}$ may be dominated by the fine-structure (FS) separation of 
the lower term, in which case the above approximation is justified if 
$\omega_{\rm FS}/\omega_k\ll 1$. For example, in the case of the 
H$\alpha$ line, $\omega_{\rm FS}/\omega_k\approx 2.4\times 10^{-5}$.

To conclude this section, we formally demonstrate that the illumination 
of the atom by a spectrally flat radiation $S_j(\omega_0,\bm{\hat{k}})$
produces a line profile that corresponds to the one obtained in the 
limit of complete redistribution (see, e.g., \citealt{He54}, \S20,
or \citealt{Sa67}, \S2.6).
In order to see this, we consider again the emissivity term associated 
with the coherent scattering of radiation (cf.\
Equations~(\ref{eq:2emiss.1}) and (\ref{eq:Jtens})), along with the 
integral norm of 
Equation~(\ref{eq:integral2}),
\begin{eqnarray} \label{eq:coherent.flat}
\varepsilon\apx{2}_i(\omega_{k'},\bm{\hat{k}'})
&\equiv&
	\frac{4}{3}\frac{e_0^4}{\hbar^2 c^4}\,{\cal N}\omega_{k'}^4
	\sum_{ll'}\rho_{ll'}\sum_{uu'l''}
	\sum_{qq'}\sum_{pp'}(-1)^{q'+p'}\,
	(r_q)_{ul}(r_{q'})^\ast_{u'l'}
	(r_p)_{u'l''}(r_{p'})^\ast_{ul''} \nonumber \\
&&{}\times
	\sum_{KQ}\sum_{K'Q'}\Pi_{KK'}\,
	\thrj{1}{1}{K}{-q}{q'}{-Q}
	\thrj{1}{1}{K'}{-p}{p'}{-Q'}\,
	T^{K'}_{Q'}(i,\bm{\hat{k}'}) \nonumber \\
&&{}\times
	\frac{2\pi}{\epsilon_{uu'}+{\rm i}\omega_{uu'}}
	\left(
	\Phi_{l''u'}^{+k'} + \bar\Phi_{l''u}^{+k'}
	\right)
	\oint \frac{d\bm{\hat{k}}}{4\pi}\,
      \sum_{j=0}^3 T^K_Q(j,\bm{\hat{k}})\,S_j(\omega_0,\bm{\hat{k}})
\nonumber \\
\noalign{\eject}
&=& 
	\frac{1}{2\pi^2\sqrt3}\,\frac{e_0^2}{c^3}\,
	{\cal N}\omega_{k'}^4
	\sum_{uu'} \Biggl\{
	\frac{16\pi^3}{\sqrt3}\frac{e_0^2}{\hbar^2 c}\,
	\frac{1}{\epsilon_{uu'}+{\rm i}\omega_{uu'}}
	\sum_{ll'}\rho_{ll'}
	\sum_{qq'}(-1)^{q'+1}\,
	(r_q)_{ul}(r_{q'})^\ast_{u'l'} \nonumber \\
&&\kern 3.5cm\times
	\sum_{KQ}\Pi_{K}
	\thrj{1}{1}{K}{-q}{q'}{-Q}
	J^K_Q(\omega_0)
	\Biggr\} %\\
%&&{}\times
	\sum_{l''}\sum_{pp'}(-1)^{p'+1}\,
	(r_p)_{u'l''}(r_{p'})^\ast_{ul''} \nonumber \\
&&{}\times
	\sum_{K'Q'}\Pi_{K'}
	\thrj{1}{1}{K'}{-p}{p'}{-Q'}\,
	T^{K'}_{Q'}(i,\bm{\hat{k}'})
	\left(
	\Phi_{l''u'}^{+k'} + \bar\Phi_{l''u}^{+k'}
	\right)\;.\qquad (i=0,1,2,3)
\end{eqnarray}
On the other hand, using the same notation, the emissivity of the first 
order is given by
\begin{eqnarray} \label{eq:incoherent.flat}
\varepsilon\apx{1}_i(\omega_{k'},\bm{\hat{k}'})
&=& \frac{1}{2\pi^2\sqrt3}\,\frac{e_0^2}{c^3}\,
	{\cal N}\omega_{k'}^4
	\sum_{uu'}\rho_{uu'}\sum_{l''}\sum_{pp'}(-1)^{p'+1}\,
	(r_p)_{u'l''}(r_{p'})_{ul''}^* \\
&&{}\times
	\sum_{K'Q'}\Pi_{K'}\,\thrj{1}{1}{K'}{-p}{p'}{-Q'}\,
	T^{K'}_{Q'}(i,\bm{\hat{k}'})
	\left(
	\Phi_{l''u'}^{+k'} + \bar\Phi_{l''u}^{+k'}
	\right)\;.\qquad (i=0,1,2,3) \nonumber
\end{eqnarray}
We then see that Equations~(\ref{eq:coherent.flat}) and
(\ref{eq:incoherent.flat}) coincide \emph{in the case 
of spectrally flat illumination}, once we identify the quantity within 
curly brackets in Equation~(\ref{eq:coherent.flat}) 
with the upper-term density matrix $\rho_{uu'}$ that appears in
the expression of $\varepsilon\apx{1}_i(\omega_{k'},\bm{\hat{k}'})$.
In fact, this identification derives directly from the solution of the
statistical equilibrium problem for the two-term atom illuminated by
spectrally flat radiation (and neglecting stimulated emission), which 
is obtained within the lowest-order approximation of the theory of 
polarized line formation \citep{LL04}. 
In turn, \emph{this proves the validity of that former theory 
for the description of the scattering of polarized radiation in the limit 
of complete redistribution.}

It is important to remark that the above result does \emph{not} imply that 
the emitted radiation gets counted twice when we include the 
coherent-scattering term in the radiative transfer equation, as this would 
evidently violate energy conservation. We have previously noted that
the excitation of $\rho_{uu'}$ is inhibited in the limit of infinitely 
sharp lower levels, and therefore all the emitted radiation must 
come from the coherent-scattering term in that case.
%$\chi_i(\omega_{k'},\bm{\hat{k}'})$. 
Then, if the illumination of the atom is spectrally flat,
% -- so that the condition for complete redistribution is satisfied -- 
%the above demonstration shows that the coherent-scattering term fully 
%accounts for the radiation that is re-emitted after the atom has been 
%radiatively excited. 
the above demonstration shows that the scattered radiation has exactly 
the same spectral structure \emph{as if} the atom had absorbed and 
re-emitted incoherently the incident radiation \cite[cf.][Sect.~20]{He54}.

%
%In contrast, when the lifetime of the lower term for transitions 
%towards the upper term is finite (i.e., the lower levels are broadened),
%%\footnote{In our treatment, this can only happen in the 
%%case of non-radiative broadening of the atomic levels, such as in the 
%%presence of inelastic collisions.} 
%the upper levels' populations and coherences have finite values, 
%and thus both terms (\ref{eq:coherent.flat}) and 
%(\ref{eq:incoherent.flat}) must contribute to the scattered radiation. 
%However, we also must expect that the statistical-equilibrium solution for 
%$\rho_{uu'}$ will be different from the one derived in the lowest-order 
%approximation of the atom-photon interactions, since a new ``channel''
%corresponding to coherent scattering has been added to the picture. 
%In particular, \emph{in the absence of collisions}, when the lower term 
%is radiatively broadened we expect that the scattered radiation will 
%still be predominantly accounted for by the coherent-scattering term 
%(as long as $A_{ul}\gg B_{lu} J^0_0(\omega_{ul})$;
%cf.~Eq.~[\ref{eq:Jtens}]), whereas the 1st-order emissivity 
%will only describe that fraction of the scattered radiation that 
%is emitted incoherently after the excitation of the upper term.

%Finally we must observe that, for a completely self-consistent 
%demonstration of energy conservation, the contribution of stimulation 
%effects can no longer be neglected when the lower term is radiatively 
%broadened.

\section{Conclusion}
\label{sec:conclusion}

We presented a general expression of the redistribution function 
for polarized radiation in the case of a transition between two atomic 
terms $l$ and $u$, which can both be partially degenerate and polarized, 
and subject to a magnetic field of arbitrary strength. In particular,
this function applies to the description of resonance or subordinate
lines of atoms with both fine and hyperfine structures. This function is 
derived from first principles, and it represents a preliminary result 
of a diagrammatic theory of scattering polarization that we have been
developing, on and off, over the past decade. The full theory is 
still incomplete, lacking a self-consistent set of accompanying 
equations for the statistical equilibrium of the atomic 
system. Part of the problem is caused by a mild violation of the 
positivity condition for the atomic density matrix in the presence of
magnetic fields. 
%This may be caused by the loss of unitarity of the 
%$S$-matrix, following the partial resummation of the self-energy diagram 
%that leads to the ``dressing'' of the atomic propagator. 
Following a revision of 
our diagrammatic approach, inspired by certain literature dealing with 
the problem of the treatment of the initial conditions in perturbative 
master equations \cite[see, e.g.,][and references within; a very clear 
discussion of this problem can also be found in a recent paper by 
\citealt{Ko10}]{Wh08}, we have identified new terms, previously
unaccounted for, in the perturbative expansion of the 
evolution equation
for the atomic density matrix. These terms are in the process of being 
evaluated, with the perspective that they should restore the 
self-consistency of the statistical-equilibrium problem.

Because of the approximations of the theory, we cannot rigorously model 
the radiative broadening of the atomic levels due to photon 
absorption. Therefore, in the absence of collisions, the lower term 
of an atomic transition remains infinitely sharp, and the scattered 
radiation only comes from the coherent contribution to the emissivity,
Equation~(\ref{eq:2emiss}).
However, we presented the general form of the redistribution 
function, Equation~(\ref{eq:red.gen}), where the widths of the lower 
levels appear explicitly. This function can then be applied to treat 
phenomenologically the frequency redistribution of radiation in the 
presence of collisional broadening. 
%
%In fact, a completely self-consistent description of radiation scattering 
%in the case of broadened lower levels can only be attained through a 
%formalism that also takes into account the collisional problem, alongside 
%with the radiative problem that has been considered in this paper.

Despite these formal difficulties, the results we presented are 
relevant enough for the spectro-polarimetric diagnostics of the 
upper layers of the Sun's atmosphere, that we decided to propose them, 
already at this early stage, to the astrophysical community.

In this paper, we limited ourselves to provide the algebraic formulas
for the scattered radiation in a complex two-term atom with hyperfine
structure, and in the presence of a magnetic field (relevant, for 
instance, to the case of the resonant \ion{Na}{1} D-doublet, or of 
complex subordinate lines such as H$\alpha$). Numerical applications 
of this formalism will be considered in future papers.

Finally, it is important to note that the general redistribution 
function presented in this paper, Equation~(\ref{eq:red.gen}) or 
(\ref{eq:red.gen.alt}), coincides with a rarely quoted result by 
\cite{LT71}, which was derived through a completely different approach 
to the problem of resonance scattering (based on \citealt{He54}). 
More precisely, these authors provided the expression 
of a ``resonance matrix'' for scattering (see their Equation~(7.7)),
which turns out to be equivalent to
\begin{displaymath}
[(\epsilon_{uu'}+{\rm i}\omega_{uu'})
  (\epsilon_{ll'}+{\rm i}\omega_{ll'})]^{-1}\,
 {\cal R}(\Omega_u,\Omega_{u'};\Omega_l,\Omega_{l'},\Omega_{l''};
	\omega_k,\omega_{k'})\;,
\end{displaymath}
in our notation. 
This correspondence allows us to derive yet another form for the general 
redistribution function presented in this paper, which is fully
equivalent to Equations~(\ref{eq:red.gen}) and (\ref{eq:red.gen.alt}),
\begin{eqnarray} \label{eq:r_l&th}
{\cal R}(\Omega_u,\Omega_{u'};
	\Omega_l,\Omega_{l'},\Omega_{l''};
	\omega_k,\omega_{k'}) \\
%&\equiv &
%	(\epsilon_{uu'}+{\rm i}\omega_{uu'})
%	\left(
%	\Psi_{u'l',l''ul}^{-k,+k'-k} + \bar\Psi_{ul,l''u'l'}^{-k,+k'-k}
%	\right) \nonumber \\
&&\kern -1.75in
{}=
\Phi^{+k}_{lu'}\,\bar\Phi^{+k'-k}_{l''l} + 
\bar\Phi^{+k'-k}_{l''l}\,\bar\Phi^{+k'}_{l''u} +  
\bar\Phi^{+k}_{l'u}\,\bar\Phi^{+k'}_{l''u} + 
\Phi^{+k'-k}_{l''l'}\,\bar\Phi^{+k}_{l'u} + 
\Phi^{+k}_{lu'}\,\Phi^{+k'}_{l''u'} +  
\Phi^{+k'-k}_{l''l'}\,\Phi^{+k'}_{l''u'} \nonumber \\
&&\kern -1.75in
{}=
\Phi^{+k}_{lu'}\,\Phi^{+k-k'}_{ll''} + 
\Phi^{+k-k'}_{ll''}\,\Phi^{-k'}_{ul''} +  
\Phi^{-k}_{ul'}\,\Phi^{-k'}_{ul''} + 
\Phi^{+k'-k}_{l''l'}\,\Phi^{-k}_{ul'} + 
\Phi^{+k}_{lu'}\,\Phi^{+k'}_{l''u'} +  
\Phi^{+k'-k}_{l''l'}\,\Phi^{+k'}_{l''u'}\;, \nonumber
\end{eqnarray}
where in the last line we used the conjugation property of the
profiles $\Phi^{\pm k}_{ab}$ (cf.~Equation~(\ref{eq:Phi})),
\begin{equation}
\bar\Phi^{\pm k}_{ab}=\Phi^{\mp k}_{ba}\;.
\end{equation}

The equivalence between the general redistribution function presented
in this paper and the form (\ref{eq:r_l&th}) derived by \cite{LT71}
gives us additional confidence of the solidity of our results and of 
the formalism on which they are based. It is regrettable that such 
fundamental result of \cite{LT71} has essentially gone unnoticed
by the literature published over the past 
forty years on the subject of partial redistribution for polarized 
radiation. We hope that 
our paper will at least succeed in bringing once again to the 
attention of the solar polarimetry community the beauty and 
relevance, still to this day, of that seminal work.

\appendix

\section{Frequency redistribution in the laboratory reference frame}

In order to express the redistribution function 
${\cal R}(\Omega_u,\Omega_{u'};
	\Omega_l,\Omega_{l'},\Omega_{l''};
	\omega_k,\omega_{k'})$
in the reference frame of the emitting plasma (the ``laboratory''
frame), it is convenient to start from the form (\ref{eq:r_l&th}).

Let $\hat{\omega}_k$ and $\hat{\omega}_{k'}$ be the angular frequencies in the 
laboratory reference frame, respectively for the incident and scattered 
photons, corresponding to $\omega_k$ and $\omega_{k'}$ in the atomic 
rest frame. Then \citep{Hu62,Mi78}
\begin{equation} \label{eq:doppler0}
\omega_k = \hat{\omega}_k - \Delta\omega_T\,\bm{v}\cdot \bm{\hat{k}}\;,
\qquad
\omega_{k'} = \hat{\omega}_{k'} - \Delta\omega_T\,\bm{v}\cdot \bm{\hat{k}'}\;.
\end{equation} 
Here $\bm{v}$ is the velocity of the scattering atom in the laboratory
frame, expressed in units of the thermal
velocity $v_T=\sqrt{2k_{\rm B}T/M}$, where $k_{\rm B}$ is the Boltzmann 
constant, $T$ the temperature of the plasma, and $M$ the mass of the
atom.
In turn the Doppler width is given by $\Delta\omega_T=\omega_0\,v_T/c$,
where $\omega_0$ is the ``average'' frequency of the atomic transition.
In the following treatment we ignore (as it is customarily done) the
effects of the $\omega_{k'}^4$ factor in the expression of the scattering
emissivity (cf.~Equation~(\ref{eq:2emiss.1})).

In order to obtain the angle-dependent redistribution function in the 
laboratory frame, we must subtitute eqs.~(\ref{eq:doppler0}) into 
Equation~(\ref{eq:r_l&th}), and average the resulting expression 
over a Maxwellian distribution for the velocity of the scattering atom,
which in the normalized units just introduced is simply given by
\begin{equation} \label{eq:Maxwellian}
P(\bm{v})=\pi^{-3/2}\exp(-\bm{v}\cdot\bm{v})\;.
\end{equation}

We introduce first a Cartesian system
$\{\bm{\hat{n}}_i\}_{i=1,2,3}$ such that $\bm{\hat{n}}_1$ and 
$\bm{\hat{n}}_2$ are coplanar with $\bm{\hat{k}}$ and $\bm{\hat{k}'}$, 
and we indicate with $(p,q,r)$ the component vector of $\bm{v}$ 
in this system. 
We define the scattering angle $\Theta$ such that 
$\cos\Theta=\bm{\hat{k}}\cdot\bm{\hat{k}'}$, and
for convenience we introduce the shorthand notations
\begin{equation}
C_2=\cos(\Theta/2)\;,\qquad S_2=\sin(\Theta/2)\;.
\end{equation}
We then choose $\bm{\hat{n}}_1$ as the bisector of the 
scattering angle, so that \cite[e.g.,][]{Mi78}
\begin{equation} \label{eq:doppler2}
\bm{v}\cdot\bm{\hat{k}}=C_2\,p + S_2\,q\;,
\qquad
\bm{v}\cdot\bm{\hat{k}'} =C_2\,p - S_2\,q\;.
\end{equation}
%
%This choice is adequate to treat all the contributions to
%eq.~(\ref{eq:r_l&th}), with the exception of the 3rd and 5th addenda,
%for which it is more convenient to choose $\bm{\hat{n}}_1=\bm{\hat{k}}$
%instead, which gives
%%
%\begin{equation} \label{eq:doppler1}
%\bm{v}\cdot\bm{\hat{k}}=p\;, \qquad
%\bm{v}\cdot\bm{\hat{k}'}=C_1\,p + S_1\,q\;.
%\end{equation}
%%

For simplicity we assume $\epsilon_u=\epsilon_{u'}$ and 
$\epsilon_l=\epsilon_{l'}=\epsilon_{l''}$, so it is possible to
introduce only two (dimensionless) damping parameters 
$a_{u,l}=\epsilon_{u,l}/\Delta\omega_T$. We note that this is a
reasonable approximation in the case of a two-term atom.
Finally we introduce the normalized frequencies 
\begin{equation}
v_{ab}=\frac{\hat{\omega}_k-\omega_{ab}}{\Delta\omega_T}\;,
\qquad
w_{ab}=\frac{\hat{\omega}_{k'}-\omega_{ab}}{\Delta\omega_T}\;.
\end{equation}
for any two levels $a$ and $b$.

After some tedious algebraic manipulation, and assuming
$\Theta\ne0,\pi$ \citep[see][for comments about those limiting cases]{MK80},
we find
%%
%\begin{eqnarray} \label{eq:rlab}
%R(\Omega_u,\Omega_{u'};\Omega_l,\Omega_{l'},\Omega_{l''};
%	\hat{\omega}_{k},\hat{\omega}_{k'};\Theta) &=& \frac{\pi}{\Delta\omega_T^2\,S_1} \\
%&&\kern -2in\times\Biggl\{
%\int_{-\infty}^\infty dq\; {\rm e}^{-q^2}\,
%\frac{\overline{W}\bigl( (v_{u'l}-S_2\,q)/C_2,(a_u+a_l)/C_2 \bigr)}
%{a_l/S_2+{\rm i}\bigl[\frac{1}{2}(v_{u'l}-w_{u'l''})/S_2 - q\bigr]}
%\nonumber \\
%&&\kern -2in {}+
%\int_{-\infty}^\infty dq\; {\rm e}^{-q^2}\,
%\frac{W\bigl( (w_{ul''}+S_2\,q)/C_2,(a_u+a_l)/C_2 \bigr)}
%{a_l/S_2+{\rm i}\bigl[\frac{1}{2}(v_{ul}-w_{ul''})/S_2 - q\bigr]}
%\nonumber \\
%&&\kern -2in {}+
%\int_{-\infty}^\infty dp\; {\rm e}^{-p^2}\,
%\frac{W\bigl( (w_{ul''}-C_1\,p)/S_1,(a_u+a_l)/S_1 \bigr)}
%{a_u+a_l-{\rm i}(v_{ul'} - p)}
%\nonumber \\
%&&\kern -2in {}+
%\int_{-\infty}^\infty dq\; {\rm e}^{-q^2}\,
%\frac{W\bigl( (v_{ul'}-S_2\,q)/C_2,(a_u+a_l)/C_2 \bigr)}
%{a_l/S_2-{\rm i}\bigl[\frac{1}{2}(v_{ul'}-w_{ul''})/S_2 - q\bigr]}
%\nonumber \\
%&&\kern -2in {}+
%\int_{-\infty}^\infty dp\; {\rm e}^{-p^2}\,
%\frac{\overline{W}\bigl( (w_{u'l''}-C_1\,p)/S_1,(a_u+a_l)/S_1 \bigr)}
%{a_u+a_l+{\rm i}(v_{u'l} - p)}
%\nonumber \\
%&&\kern -2in {}+
%\int_{-\infty}^\infty dq\; {\rm e}^{-q^2}\,
%\frac{\overline{W}\bigl( (w_{u'l''}+S_2\,q)/C_2,(a_u+a_l)/C_2 \bigr)}
%{a_l/S_2-{\rm i}\bigl[\frac{1}{2}(v_{u'l'}-w_{u'l''})/S_2 - q\bigr]}
%\Biggr\}\;,
%\nonumber
%\end{eqnarray}
%%
%or
%
\begin{eqnarray} \label{eq:rlab}
R(\Omega_u,\Omega_{u'};\Omega_l,\Omega_{l'},\Omega_{l''};
	\hat{\omega}_{k},\hat{\omega}_{k'};\Theta) &=&
\frac{1}{2\Delta\omega_T^2\,C_2 S_2} 
\int_{-\infty}^\infty dq\; {\rm e}^{-q^2} \\
&&\kern -2.2in\times\Biggl\{
\biggl[
\frac{W\bigl( (v_{ul'}-S_2\,q)/C_2,(a_u+a_l)/C_2 \bigr)}
{a_l/S_2-{\rm i}\bigl[\frac{1}{2}(v_{ul'}-w_{ul''})/S_2 - q\bigr]}
+
\frac{\overline{W}\bigl( (v_{u'l}-S_2\,q)/C_2,(a_u+a_l)/C_2 \bigr)}
{a_l/S_2+{\rm i}\bigl[\frac{1}{2}(v_{u'l}-w_{u'l''})/S_2 - q\bigr]}
\biggr] \nonumber \\
&&\kern -2.2in \,\,{}+
\biggl[
\frac{W\bigl( (w_{ul''}+S_2\,q)/C_2,(a_u+a_l)/C_2 \bigr)}
{a_l/S_2+{\rm i}\bigl[\frac{1}{2}(v_{ul}-w_{ul''})/S_2 - q\bigr]}
+
\frac{\overline{W}\bigl( (w_{u'l''}+S_2\,q)/C_2,(a_u+a_l)/C_2 \bigr)}
{a_l/S_2-{\rm i}\bigl[\frac{1}{2}(v_{u'l'}-w_{u'l''})/S_2 - q\bigr]}
\biggr] \nonumber \\
&&\kern -2.2in \,\,{}+
\biggl[
\frac{W\bigl( (v_{ul'}-S_2\,q)/C_2,(a_u+a_l)/C_2 \bigr)
	- W\bigl( (w_{ul''}+S_2\,q)/C_2,(a_u+a_l)/C_2 \bigr)}
{{\rm i}\bigl[\frac{1}{2}(v_{ul'}-w_{ul''})/S_2 - q\bigr]} \nonumber \\
&&\kern -2.2in \hphantom{\frac{1}{C_2}\biggl[}\,\,{}-
\frac{\overline{W}\bigl( (v_{u'l}-S_2\,q)/C_2,(a_u+a_l)/C_2 \bigr)
	-\overline{W}\bigl( (w_{u'l''}+S_2\,q)/C_2,(a_u+a_l)/C_2 \bigr)}
{{\rm i}\bigl[\frac{1}{2}(v_{u'l}-w_{u'l''})/S_2 - q\bigr]}
\biggr] \Biggr\}\;,
\nonumber
\end{eqnarray}
where
\begin{equation}
W(v,a)=\frac{1}{\pi}\int_{-\infty}^\infty dp\;
\frac{{\rm e}^{-p^2}}{a+{\rm i}(p-v)}=H(v, a)+{\rm i}\,L(v, a)\;,
\end{equation}
$H(v,a)$ and $L(v,a)$ being the Voigt and Faraday-Voigt functions
\cite[e.g.,][]{LL04}, and we have indicated with $\overline{W}$ the
complex conjugate of $W$.
In the simplest case of a two-level atom with degenerate sublevels,
Equation~(\ref{eq:rlab}) properly reduces to
\begin{eqnarray}
R(\Omega_u;\Omega_l;
	\hat{\omega}_{k},\hat{\omega}_{k'};\Theta) 
&=& \frac{1}{\Delta\omega_T^2\,C_2 S_2^2}\,a_l
\int_{-\infty}^\infty dq\;\frac{{\rm e}^{-q^2}}
{(a_l/S_2)^2+\bigl[\frac{1}{2}(v_{ul}-w_{ul})/S_2 - q\bigr]^2} \\
&&\kern -1.6in\times\Biggl\{
\Bigl[
H\bigl( (v_{ul}-S_2\,q)/C_2,(a_u+a_l)/C_2 \bigr) +
H\bigl( (w_{ul}+S_2\,q)/C_2,(a_u+a_l)/C_2 \bigr)
\Bigr]
\nonumber \\
&&\kern -1.6in \,\,{}+ 
\frac{a_l/S_2}
	{\frac{1}{2}(v_{ul}-w_{ul})/S_2 - q}\,
\Bigl[
L\bigl( (v_{ul}-S_2\,q)/C_2,(a_u+a_l)/C_2 \bigr) -
L\bigl( (w_{ul}+S_2\,q)/C_2,(a_u+a_l)/C_2 \bigr) \Bigr]
\Biggr\}\;, \nonumber
\end{eqnarray}
which corresponds to Equation~(21) of \citeauthor{MN86}
\citep[\citeyear{MN86}; see also ][]{He81}.

\end{document}